\documentclass[a4paper,12pt,onecolumn,draftclsnofoot]{IEEEtran}
\usepackage{amssymb}
\usepackage{tabularx}
\usepackage{graphicx}
\usepackage{caption}
\usepackage{subcaption}
\usepackage[tbtags]{amsmath}
\usepackage{amsfonts}
\usepackage{mathrsfs}
\usepackage{multirow}
\usepackage[english]{babel}
\usepackage{cite}
\usepackage{url}
\usepackage{hyperref}
\hypersetup{
  colorlinks=true,
  linkcolor=red,
  linktoc=all,
  citecolor=blue,
  bookmarksnumbered=true
}
\usepackage{cleveref}
\usepackage{units}
\usepackage[section]{placeins}
\usepackage[colorinlistoftodos]{todonotes}

\IEEEoverridecommandlockouts

\begin{document}
\title{Modelling the Second Covid-19 Wave in  Mumbai}
\author{
\IEEEauthorblockN{
TIFR Covid-19 City-Scale Simulation Team \\
Sandeep Juneja,
Daksh Mittal\footnote{Daksh Mittal is supported through the IDFC Institute} \\
}
{May 2, 2021} \\
}
\maketitle

{\let\thefootnote\relax
  \footnotetext{
    Source code available at \url{https://github.com/dasarpmar/epidemics-simulator-mumbai/releases/tag/v5.0}
  }
}
\vspace*{-2.5cm}
\section{Summary}
India has been hit by a huge second wave of Covid-19 that started in mid-February 2021. 
The wave first arose in the state of Maharashtra,  and within it, Mumbai was amongst the first cities to see the
increase.
 Was this wave caused by increased laxity in the population around December and  January? Or by the opening up
 of the economy and public transportation on February 1, particularly in Mumbai?
 Did variants of the SARS-CoV-2 virus that have emerged in Maharashtra and other parts of India
 play a role in the sharpness with which the virus has spread? 
 Very likely all three played a role, and especially the variants.
While the  significance  of the role of variants remains to be confirmed by researchers,
 it appears that the double mutant variant B.1.617 \cite{double_mutant} may have played a key role
 in Maharashtra.  Unfortunately, we currently lack the data  
 to firmly establish 
 infectiousness, prevalence and virulence of  B.1.617, or that of other variants that may be active, 
compared to the strains that were prevalent in Mumbai last year.

Fortunately, vaccinations have started in India since mid-January 
and are being administered with some urgency in Mumbai since
mid-March and are likely to reduce infections and fatalities in the coming months.   
 
 In this report, we use  and enhance the IISc-TIFR city simulator \cite{City_Simulator_IISc_TIFR_2020} to computationally study
 the second wave in Mumbai. We  build upon our earlier analysis \cite{October_report_2020}, where
 projections were made from November 2020 onwards. 
 Given the inherent uncertainty in the role of the variants, and other important infection
 spread determining factors,
 we use our simulator to conduct an extensive scenario analysis - we play out
 many plausible scenarios through varying economic activity,
 reinfection levels, population compliance, infectiveness, prevalence
 and lethality of  the possible variant strains, and infection spread via local trains  
 to arrive at those
 that may better explain the second wave fatality numbers.

While  realistically  there may be multiple variants operating in Mumbai,
in our scenario analysis, to maintain simplicity,  we assume that there is a single variant.
 
We observe and highlight certain outcomes that occur across all the scenarios considered.
This robustness suggests that they are likely to be accurate 
projections for Mumbai. These include:
\begin{enumerate}
\item
 The Mumbai second wave of fatalities will likely peak around the first week of May
 (First reported by us on March 31 \cite{Tweet_March_31}).
Again,  this appears largely invariant under many plausible scenarios.
 Since cases typically lead fatalities by two to three weeks, 
 this in turn suggests that the cases will peak a few weeks before first of May, as indeed appears to be the case. 
\item
In the scenarios that loosely match 
 Mumbai's vaccination drive (age based vaccinations of the order of 15 to 20 lac new 
people a month; 75\% efficacy), the fatalities reduce to Jan. and Feb. levels by June 1. 
This again suggests that the reported cases will substantially decrease in Mumbai by early to mid May (reported
by us 
on April 15 \cite{Tweet_April_15}).
We also observe that if the extensive  vaccination 
drive  continues,  and the vaccines show the promised
efficacy (of around 75\%), without the virus escaping the vaccine immunity, 
Mumbai may be in a position to open schools by July 1 or soon thereafter.
See \cite{Lancet-India-Schools} for recommendations for opening schools in India.
Of course, as in any forecasting exercise, further off the horizon, 
 more error prone are the projections. So closer to July would be the right time to make this evaluation. 
\end{enumerate}

\begin{figure}
      \centering
     \includegraphics[width=\linewidth]{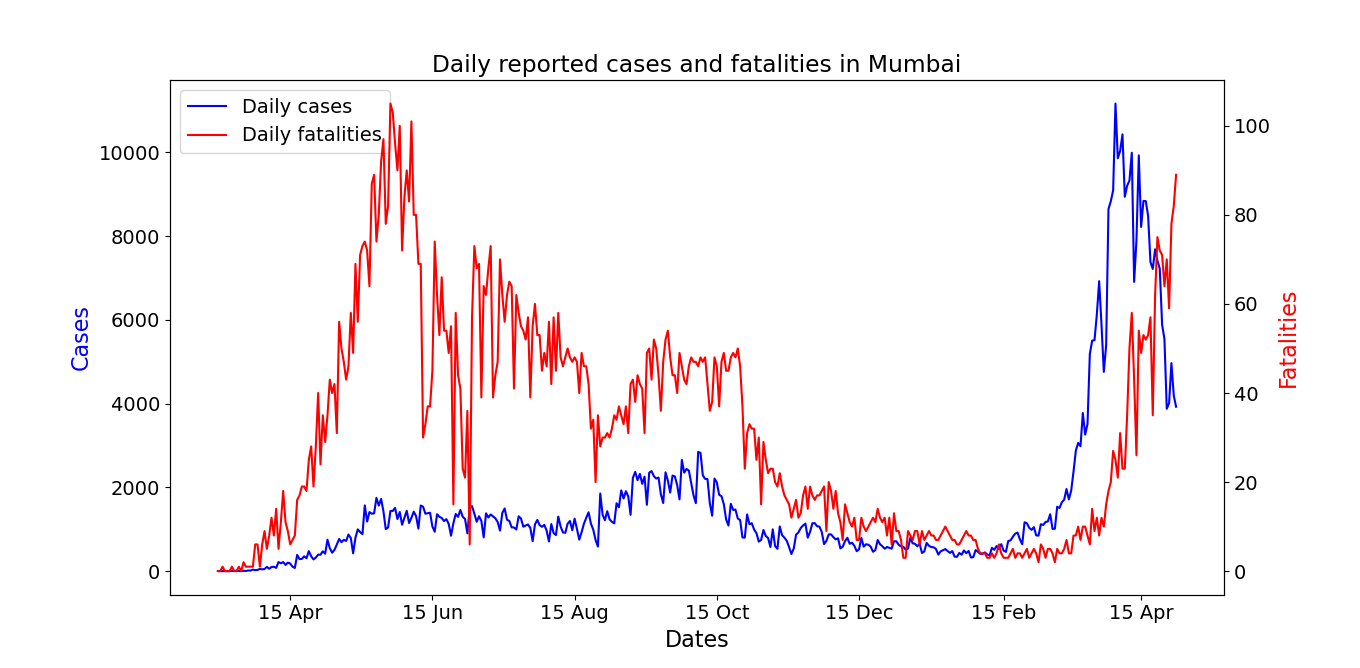}
      \caption{Daily reported cases and fatalities in Mumbai. Fatalities
      start rising around March 15 signalling a second wave.
      The rise becomes particularly steep around April 1. Cases start rising around
      mid-February signaling the second wave. They show a steep increase starting mid to late March. (\scriptsize{Source : BMC Dashboard}\normalsize{)} } \label{cases_deaths_mumbai}
  \end{figure}

\begin{figure}
      \centering
     \includegraphics[width=\linewidth]{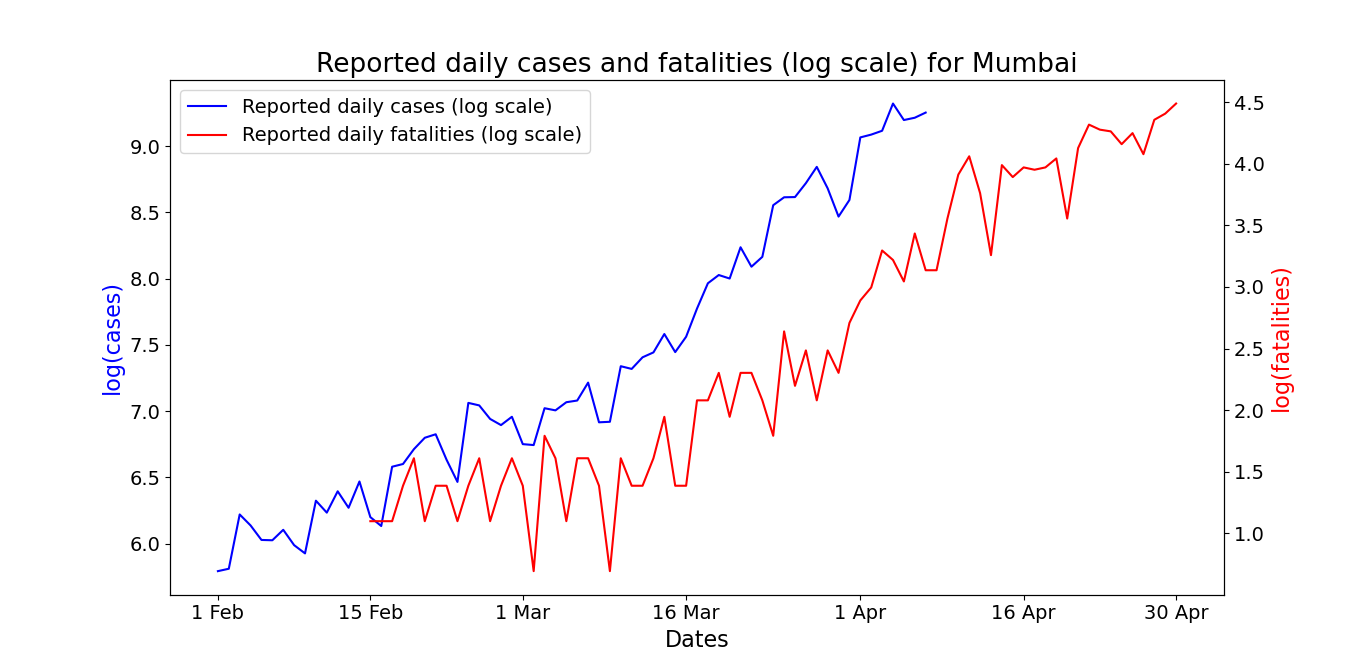}
      \caption{Daily reported cases and fatalities (log scale) in Mumbai. (\scriptsize{Source : BMC Dashboard}\normalsize{)} } \label{cases_deaths_log_scale}
  \end{figure}

During the second wave, the observed fatalities were low in February and mid-March and saw a {\em phase change}
or a steep increase in the growth rate 
after around late March  (see Fig \ref{cases_deaths_mumbai} and  Fig \ref{cases_deaths_log_scale}).
We conduct extensive experiments to replicate this observed sharp {\em convexity}. This is not an easy phenomena
to replicate, and we find that explanations such as increased laxity
in the population, increased reinfections, increased intensity of infections in Mumbai transportation,
increased lethality in the virus, or a combination amongst them, generally do a poor job of matching this pattern.

We find that the most likely
explanation is presence of small amount of extremely infective variant
on February 1  that grows rapidly thereafter and becomes a dominant strain 
by Mid-March. The scenario
where the variant is 2.5 times more infective than the dominant strain last year, and
accounts for  2.5\% of the infected population on Feb 1,  appears to match the data well. 
This variant  increases
to over 60\% by mid March. 
Another scenario  where the variant is 2 times more infective and the compliance
is less than the earlier scenario, also explains the data reasonably well.
Given the large uncertainty in model parameters, the above  specific numbers
may be far off from the truth. However,
the observation that Mumbai had highly infective variants that grew to have substantial
presence in March is likely to be true. From a prescriptive view,
this points to an urgent need for extensive and continuous genome sequencing 
to establish existence and prevalence of different virus strains in Mumbai and in India, as they 
evolve over time. This may help better manage infections that
may result from new strains in future.

\begin{figure}
      \centering
     \includegraphics[width=\linewidth]{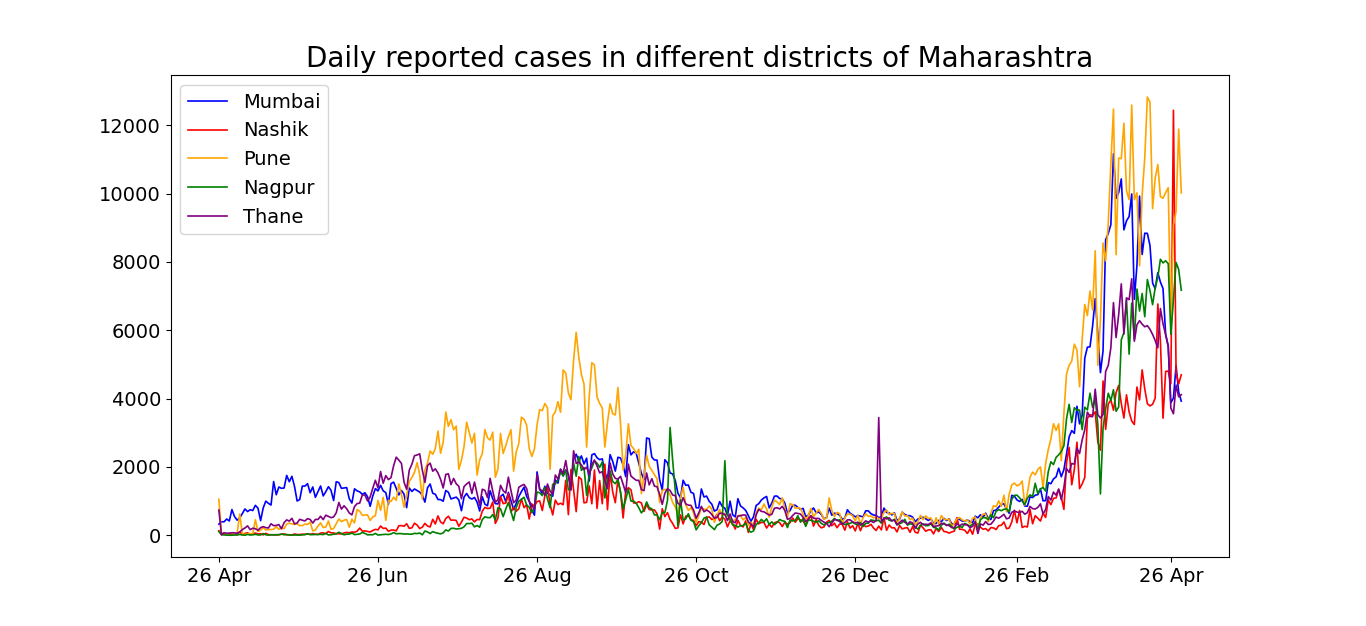}
      \caption{Daily reported cases in some of the larger  districts of Maharashtra.
      While these districts were somewhat synchronised in the August-September wave,
      they seem much more so March onwards this year.
      (\scriptsize{Source : \url{https://api.covid19india.org/}}\normalsize{)}} \label{all_districts_maharashtra_cases}
  \end{figure}

Since, other larger districts of Maharashtra such as Pune, Thane, Nashik and Nagpur are closely synchronised with Mumbai
in this second wave in terms of reported cases (see Fig \ref{all_districts_maharashtra_cases}) and other regions in the state
typically lag Mumbai, this suggests that much of Maharashtra may see a similar trend to Mumbai with 
a small lag. 

As is well known,  $R_0$ denotes the expected number of individuals a single randomly selected exposed person infects
 in a city where everyone else is susceptible.  It captures the infectivity of the virus.
Later in Section~\ref{section:R_0}  we also discuss the value of $R_0$  for the virus strain present last year  as
per our model, and its value under more infectious variants.

Given that early this year, none of the models anticipated the severe second wave that we see in India, the above claims
should be accepted with caution.  The key caveats that may affect our projection and 
to watch out for include
 a large proportion of population infected last year becoming susceptible
 to severe reinfections. The virus showing significant capacity 
 to escape the vaccine induced immunity. 
 Rise of new strains that may be more infectious and lethal.
 Of course, the city population needs to continue to guard against 
 laxity in the distancing precautions. Furthermore, while we have conducted extensive time demanding simulations,
 due to time paucity, we have not been comprehensive. Nor have we been able to implement more sophisticated
 stochastic optimization techniques. We cannot entirely rule out
 scenarios where reinfections carefully combined with population laxity may explain
 most of the observed fatality curve, even though that  appears unlikely.

\section{Analyzing the second wave through scenario analysis}

First case in Mumbai was detected on 11 March 2020.
By February 9, 2021 when the reported cases started rising and the second wave started, Mumbai had
seen 313,419 reported cases. During the ongoing second wave, 
these have sharply increased to 649,395 as of April 30, 2021 . See Fig \ref{cases_deaths_mumbai},
While the reported cases never exceeded  1,500 a day
up till July 2020 due to low testing, they
remained close to or below 2,500 between mid September and October during the minor wave  that
started in  the wake of Ganpati Festival related crowding in late August 2020. 
In comparison, the reported cases have been over 8,000 for most of the days since April 1, 2021. 
Thus, the cases are dramatically higher this time around compared to any time in 2020,
a fact that agrees with the  all India trend.

The Covid-19 reported fatalities in Mumbai (Fig \ref{cases_deaths_mumbai}) on the other hand have increased less aggressively. 
They equalled 11,396 on Feb. 9, 2021 and have increased to
13,161 as of April 30, 2021. While the data suggests that the cases have peaked
for Mumbai around second week of April, the fatalities are likely to continue to rise and peak in early May.
For perspective on rising fatality numbers, note that  
 In January, February 
and till mid-March, daily fatalities were essentially in single digits. Since then as a result of the second wave, they
have been rising to over 60 a day during April 21-25.

\begin{figure}
      \centering
     \includegraphics[width=\linewidth]{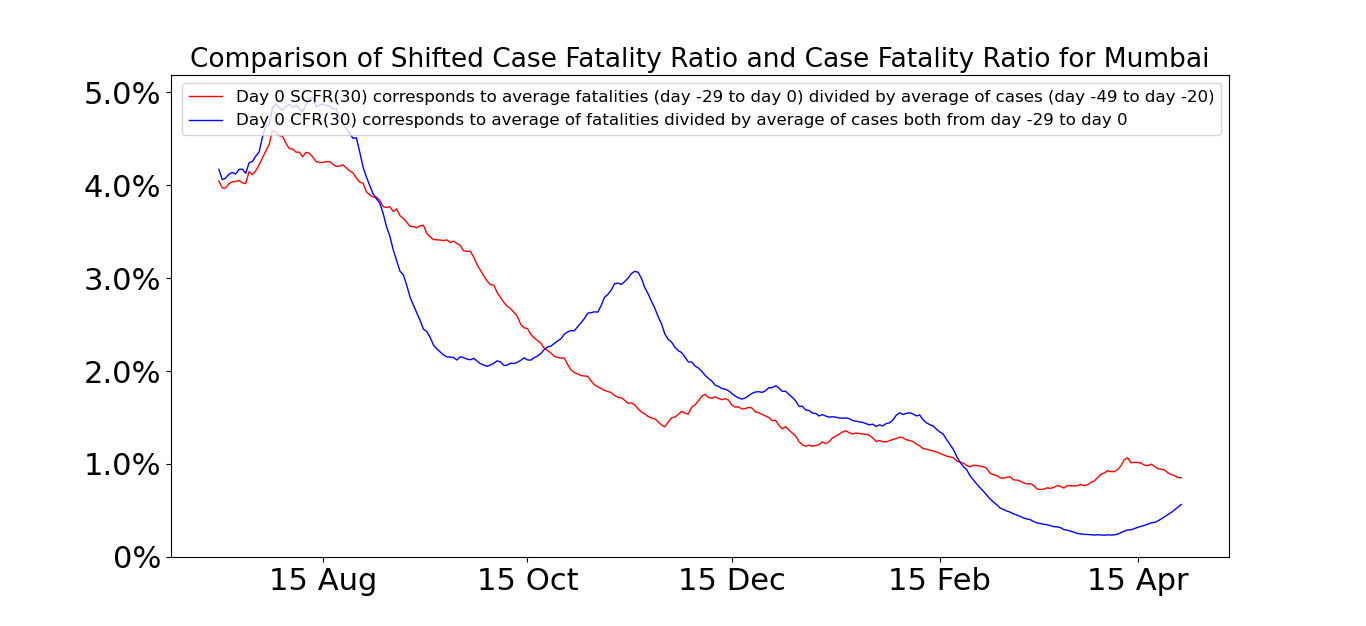}
      \caption{Comparison of Shifted Case Fatality Rate  and Case Fatality Rate for Mumbai.
      SCFR appears to be a much better measure of disease severity. (\scriptsize{Source : BMC Dashboard}\normalsize{)} } \label{CFR_SCFR_MUMBAI}

  \end{figure}

{\noindent \bf
Case Fatality Rate (CFR):} 
Could the steep increase in fatalities seen in April be largely due to avoidable deaths 
due to stretched medical infrastructure? Avoidable deaths 
mean deaths that could have been avoided if patient had access to  reasonable medical care.
This appears unlikely as the cases showed a similar steep increase mid-March onwards.
In Fig \ref{CFR_SCFR_MUMBAI}, we shed further light on  the relationship between cases and fatalities in Mumbai.
CFR is a popular measure of severity of disease. It equals  observed number of fatalities divided by 
observed number of cases during a specified time window. This can be a poor
measure when the cases are rising as fatalities typically lag cases by close to 3 weeks.
As per \cite{verity2020estimates}, the average  time between a person getting exposed
and the person not surviving is around 28 days.
Also, a person typically develops symptoms around 5 to 6 days after getting exposed. It is then reasonable
to assume that an exposed person reports positive around 8-9 days later.
With these numbers in mind, in Fig \ref{CFR_SCFR_MUMBAI}, we also report the Shifted Case Fatality Rate (SCFR) measured over a thirty day window where
the cases in the denominator are an average 20 days earlier than the fatalities in the numerator.
Thirty  days are chosen to be large enough to provide a stable estimator, while small enough
to capture time varying trends.
For Mumbai we see that while this number was over  3\% in September 2020, 
it is currently around 1\%.  The figure also illustrates that  SCFR may be a better measure
of disease severity compared to CFR that behaves poorly due to the lag between cases and fatalities. 
The fact that SCFR is taking reasonably small values suggests that Mumbai has not seen many avoidable deaths.

Apart from increased testing (see Fig \ref{tests_positivity_data_mumbai}),
 key factor that explains the discrepancy in SCFR in September 2020 and April 2021
may be that the population segment getting exposed now is more likely to get tested compared to some months ago.
 Other possible reasons include
improved medical care, 
presence of  more infectious strain (s) that maybe less deadly,
and presence of mild reinfections that get reported as cases but 
lead to reduced fatalities compared to the first time infections. 
As an interesting digression, in the appendix (section \ref{section:append}), we show SCFR graphs for other districts in Maharashtra, for India,
as well as for some of the states with the largest number of people infected in the 
second wave.  The current SCFR of over 6\% for Delhi and UP is particularly worrying.
In addition to suggesting significant reporting issues,
this also suggests presence of a more fatal strain (perhaps the UK Variant), and of 
many avoidable fatalities. The SCFR for India is around 2.5\%, while interestingly, the CFR is less
than 1\%.

\bigskip

{\noindent \bf Second Wave:} An important  aim of this report is to develop insights into the second wave in Mumbai
through playing out various scenarios and see the ones that better explain the Mumbai
fatality data from February to April 2021, with its peculiar late escalation. As in our previous reports,
we avoid trying to match observed cases because they are much less predictable - they rely on administrations' testing policy that can vary with time and location. The observed cases also rely heavily on the segment of population getting tested. For instance,
a slum dweller, for the same symptoms, is much less likely to get tested compared to
someone living in a high rise.

\begin{figure}
      \centering
     \includegraphics[width=\linewidth]{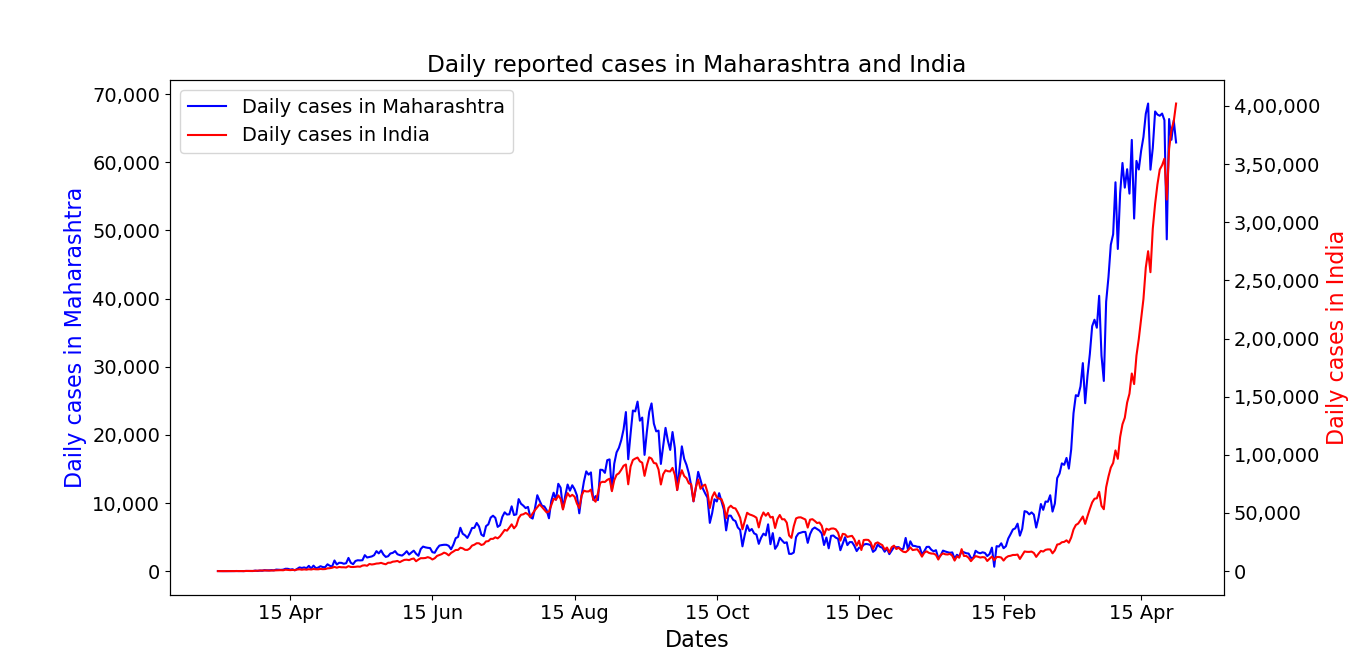}
      \caption{Daily reported cases in Maharashtra and India. The cases begin to rise
      from mid-February in Maharashtra. While the cases in India begin to rise
      from mid-March. (\scriptsize{Source : \url{https://api.covid19india.org/}}\normalsize{)}} \label{Daily_cases_MH_INDIA}

  \end{figure}

\begin{figure}
      \centering
     \includegraphics[width=\linewidth]{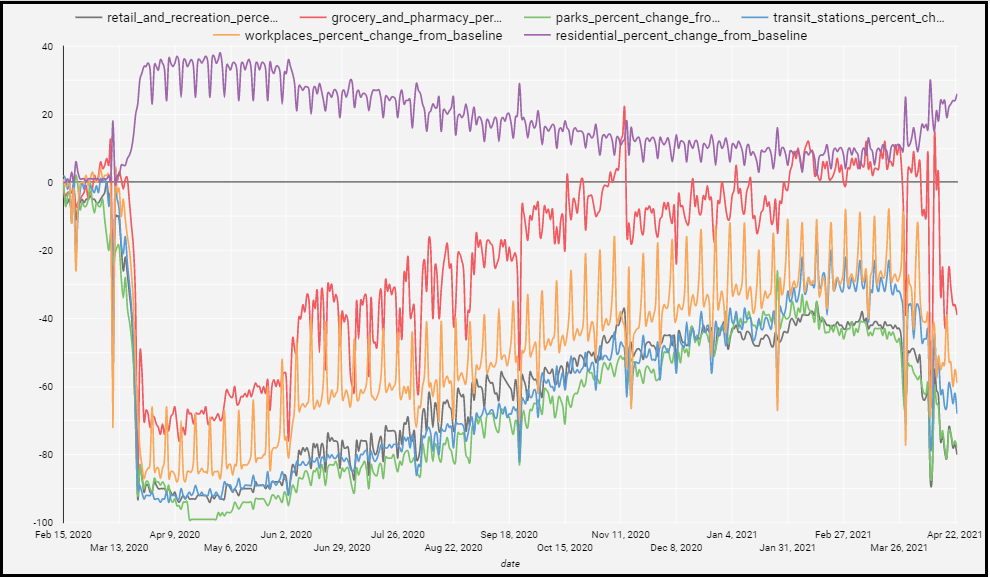}
      \caption{Google Mobility Report for Mumbai. This graph shows the percentage change in mobility from the base line in different categories.  After a steep fall in April last year,  a steady increase is seen across categories  starting April 2020 up till January 2021. Thereafter,
      the mobility is somewhat stable till March 2021, and again reduces in April 2021. (\scriptsize{Source : \url{https://datastudio.google.com/u/0/reporting/a529e043-e2b9-4e6f-86c6-ec99a5d7b9a4/page/yY2MB?s=ho2bve3abdM}}\normalsize{)}
      %Red line is for Grocery an Pharmacy related mobility, Orange line corresponds to workplace related mobility,  Green line corresponds to parks related mobility, Grey line is for retail and recreation related mobility, Blue line is for transit stations related and purple for residential related mobility. 
      } \label{google_mobility_report_mumbai}
  \end{figure}

\begin{figure}
      \centering
     \includegraphics[width=\linewidth]{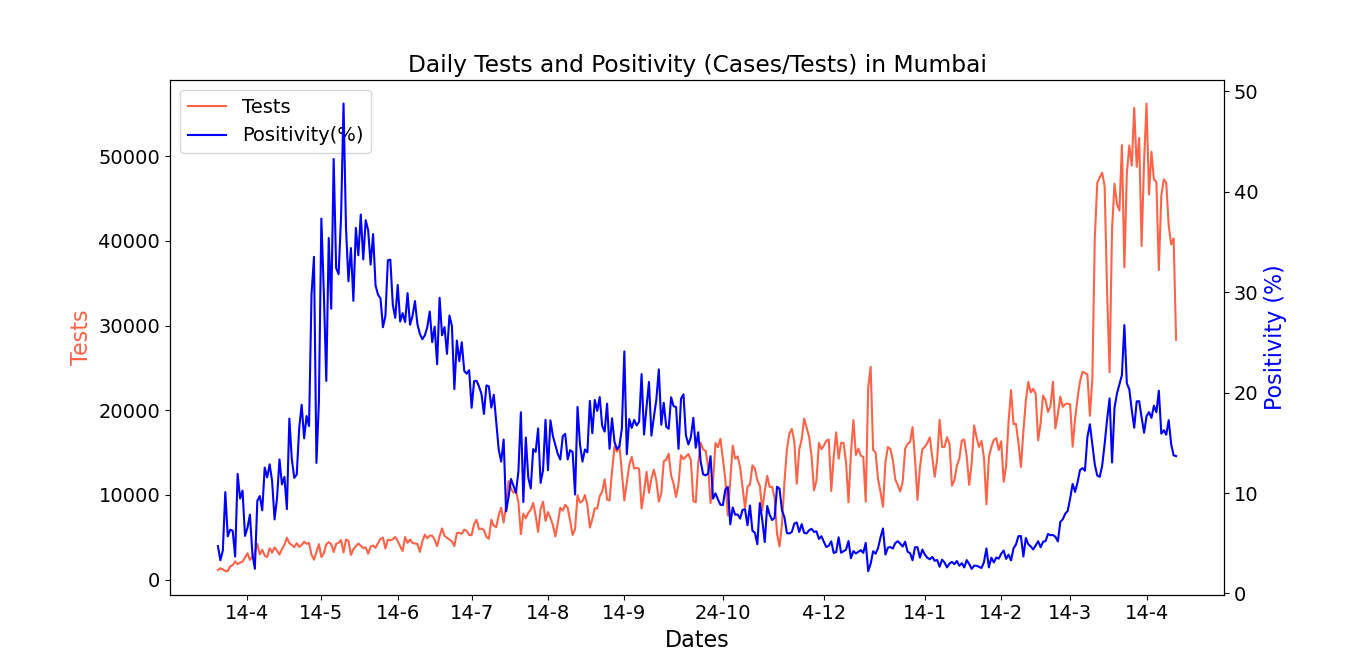}
      \caption{Daily tests in Mumbai and the case positivity. Testing is much higher in
      April 2021 compared to August and September 2020, although the positivity is similar. (\scriptsize{Source : BMC Dashboard}\normalsize{)}
      } \label{tests_positivity_data_mumbai}
  \end{figure}

Before we conduct scenario analysis, lets understand the observed fatality numbers a little better: 
Recall that 
by late January 2021, the reported Covid-19 cases in India, in Maharashtra (see Fig \ref{Daily_cases_MH_INDIA}), and in Mumbai (see Fig \ref{cases_deaths_mumbai}), were well
below their peak from last year.
Earlier Serosurveys in Mumbai \cite{malani2020serosurvey}
had shown a high prevalence of the disease especially in the dense slum areas, and it was expected that if the city fully opened by November, then by  January
it would  approach some form of herd-immunity \cite{October_report_2020}.  Serosurveys in other metropolitan cities also
showed high prevalence, and even all of India was seen to have a significant proportion of population
already exposed to the disease, see \cite{delhi_sero} and \cite{icmr_sero}.
The general 
 mood was that, given the relatively high prevalence,
the young and largely rural population
that may have some form of prior immunity,  the pandemic was over in India.  This reflected in the increased population 
mobility, increased social gatherings and increased laxness in observing social distancing. 
The Google Mobility Report \cite{google_mobility_report} in Fig. \ref{google_mobility_report_mumbai} underscores this increase. 

  %  \begin{figure}
  %    \centering
 %    \includegraphics[width=\linewidth]{april_plots/tests_positivity_BMC.PNG}
%      \caption{Daily Tests (RT-PCR and rapid Antigen) in Mumbai and the case positivity. From 
%      the MCGM Dashboard} \label{tests_positivity_BMC}
 % \end{figure}

\begin{figure}
      \centering
     \includegraphics[width=\linewidth]{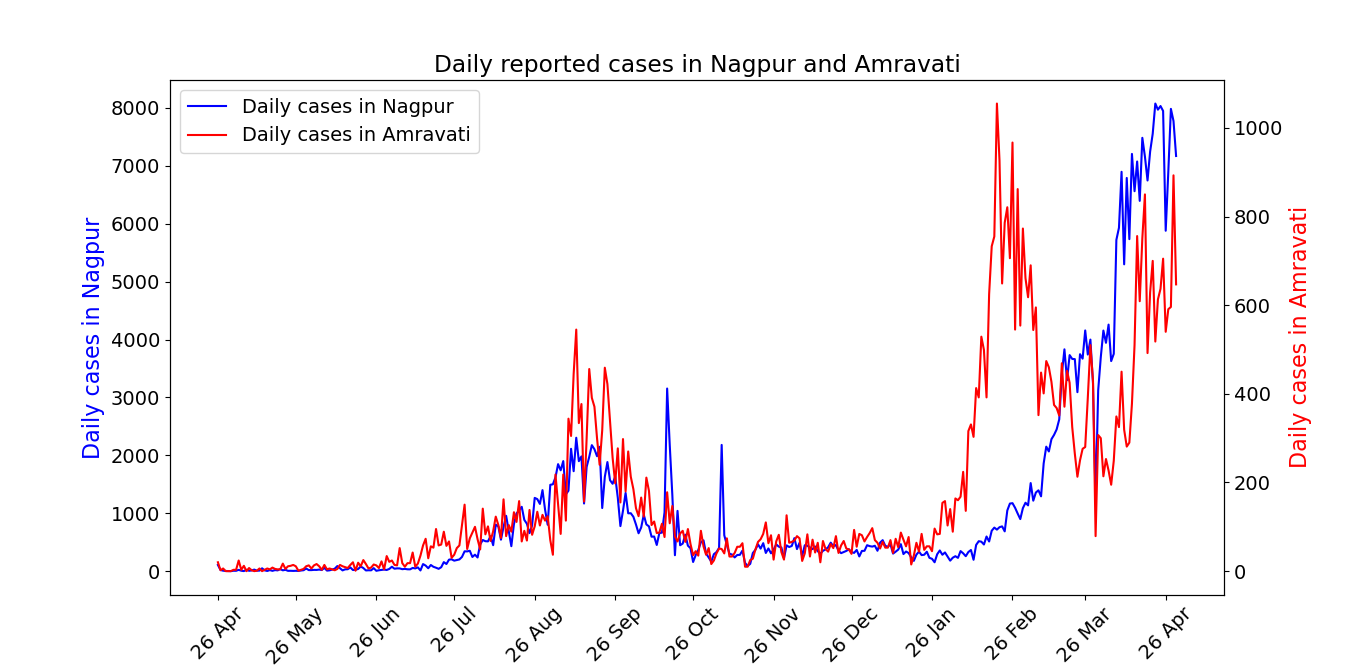}
      \caption{Daily reported cases in Nagpur and Amravati. Cases in Nagpur show an increasing trend from early February 2021. While the cases in Amravati show an increasing trend starting  January 26, 2021. (\scriptsize{Source : \url{https://api.covid19india.org/}}\normalsize{)}
      } \label{Daily_cases_nagpur_amravati}
  \end{figure}

Around late January, cases started to increase in Amravati and Nagpur in Maharashtra (see Fig \ref{Daily_cases_nagpur_amravati}).
The city of Mumbai substantially opened up on February 1 \cite{Feb_1_opening}, where 
the restrictions on the people travelling by local trains,
the lifeline of the city in normal times, were reduced. Initially, from February 9 onwards, the cases in Mumbai 
increased at a slow rate (see Fig \ref{cases_deaths_mumbai}), however, by early-March there appeared to be a {\em phase change} 
and the rate of growth increased substantially.
Similarly, the observed fatalities were low in February and mid-March and saw a steep increase in the growth rate 
after around late March (see Fig \ref{cases_deaths_mumbai}).

As we discuss later in Section \ref{section:oct_res}, projections from our October report \cite{October_report_2020} match the observed fatality graph 
quite well till December with some underestimation in January and February  (with January opening in Figure 11 shifted by a month
to roughly depict the Feb. opening). Recall that in Mumbai, restrictions on train travel 
was substantially relaxed on Feb. 1. The report was written under the assumption
that the population compliance to social distancing remains unchanged through
early 2021.   With the adjustment for population laxity in December and January,
the projected fatality data matches the observations reasonably well till early March. However,
it  completely misses the sharp rise in fatality numbers thereafter.

In this report, we conduct an extensive scenario analysis to try and explain the much larger fatalities observed from mid-March to end 
April.  We fix a base case that appears to match the fatality data well, and consider many scenarios
that are perturbations to the base case.
Our key conclusions - that the Mumbai second wave of fatalities will likely peak around the first week of May,
and that the fatalities reduce to Jan. and Feb. levels by June 1 can be seen to roughly hold
across all the scenarios that we consider.

\subsection{Base case}

{\noindent \bf Mobility:}
As in the January opening scenario of \cite{October_report_2020}, 
we assume that the mobility of the Mumbai population is at 
50\% of the pre-covid times in the months from September to December,
and extend this level to January as well.
This increases to 65\% on February 1 when
the city further opened up \cite{Feb_1_opening} and stays at that level  till the `Break the Chain' lockdown on
April 15, 2021 (see \cite{lockdown_15_april}). 
To see that this approximation is in the correct ball park, we observed that 
the data on average passenger counts travelling daily 
by buses and local trains between January 13 to January 31, 2021 equals  4.7 million,
while this average between February 1 to February 14, 2021 equals 5.8 million (source: MCGM).
Given that in normal times travel by train is around 8 million passenger counts daily \cite{mumbai_local}, and by BEST buses, it is order 
2 million \cite{mumbai_best_buses}, are selected mobility percentages appear reasonable. 
 These percentages are also in rough consonance with the  Google Mobility Report 
in Figure~\ref{google_mobility_report_mumbai}. 
One could fine tune these  numbers and also account for slight slowing down in late March and early April in
the view of rising cases, but this is unlikely to impact the model fatality profile significantly. 

Maharashtra government imposed the `Break the Chain'  lockdown restrictions
from April 15 to May 1, that were later extended to May 15 (see \cite{lockdown_1_may}).
Google Mobility Report suggests that mobility in late April was similar and slightly higher
than in June 2020. We impose
restrictions in our model similar to those in June 2020, with fraction travelling to work
increased from 15\% in June 2020 to 20\% during the new lockdown.  

\bigskip

{\noindent \bf Compliance:}
In our October report, we had assumed that compliance levels was 60\% in non-slums (NS)  and 40\%  (S)
in slums throughout the period (except for the festival periods where compliance is reduced to 40\% NS and 20\%  S). 
The reduction in cases and fatalities likely led to increased laxity in the population
in December and January.
The increased
mobility as seen in Google Mobility Report for retail and recreation, workplaces, transit stations, parks and  groceries suggests that population mixing had increased  and people likely became more
lax in social distancing, wearing masks, etc. in December 2020 and January 2021.
We account for this laxity by changing the compliance from 60\% NS and 40 \% S to  
to 50\% NS and 30\% S in December and 40\%  NS and 20\%  S in January.
In February, till the 18 th, the  compliance is assumed to be 20\% NS and 10\%  S but from February 19 onwards, it is increased to 40\% NS and 20\% S till April 14. This increase in compliance was based on the order from MCGM for compulsory masks \cite{18_feb_bmc_order}. During the lockdown (April 15 - May 15), the compliance is assumed to be 60\% NS and 40\% S.
\bigskip

{\noindent \bf Reinfection level:}
While some cases of reinfection have been reported in the literature,
the phenomena does appear to be limited so far. 
Further, if the reinfections are mostly mild, they have little 
affect on the fatality numbers. We keep the reinfection  level to zero 
in the base case. Other levels do not improve the fit to data.
Later, we conduct extensive perturbations of 
reinfection levels to assess their impact.

\bigskip

{\noindent \bf Variants:}
We assume that there exists a single variant that accounts for 2.5\% of all the infected population
on Feb. 1 in our model. These are randomly chosen amongst all the infected 
on Feb 1. Further we assume that this variant is 2.5 times more infected 
compared to the original strain.
Technically, this means that the transmission rates
in our simulation (see \cite{City_Simulator_IISc_TIFR_2020}) corresponding to transmission at home,
at workplaces, in communities and in local trains
is increased by 2.5 when an infected person, infected by the new variant, encounters
a susceptible one.  This is the transmission rate
assigned to all who are thereafter infected by the new variant.

\bigskip

{\noindent \bf Vaccination:}
The elderly (above 60 years of age) were given the first vaccination dose in Mumbai from March 1 onwards.
Vaccines were made available to 45 years and older from April 1.
The vaccine starts to provide immunity many weeks later. In the month of March around 5 lac people were vaccinated.
Since then the number is closer to 18 lac a month and is expected to become higher provided the supply continues.
In our model, we assume that vaccine once administered is immediately effective. Further, to capture vaccine's efficacy of around 75\%,
we assume that 75\% of the people vaccinated have complete protection, while randomly chosen
25\% of the vaccinated get no protection.  The specific vaccination schedule that we implement
in our model is:
\begin{itemize}
    \item 
 40\% (appx. 5 lakh) people above the age of 60 years are vaccinated in month of April (quarter of these numbers
 are vaccinated during the month at four equally spaced instances).
 \item 
 15 lakh people above the age of 45 years (this includes 40\% above age of 60) are vaccinated in month of May for the first time (again 
 four times each month).
 \item 
 20 lakh people above the age of 18 years are vaccinated for the first time in each month from May to July. ( four times each month).
\end{itemize} 
\bigskip

{\noindent \bf Impact of trains:}
In our earlier report  \cite{IISc_TIFR_2020_Mumbai_Report2}  we propose a methodology 
to model infection spread in local trains. The train transmission rate ($\beta_T$)
based on some reasonable approximations
corresponding to contact rates is set to 0.4 times $\beta_H$, where $\beta_H$ denotes the transmission rate at home . We 
continue with this setting in the base case.

\bigskip
{\noindent \bf Variant virulence:}
In the base case, we keep the virulence  at the level of the original variant. 

\bigskip
{\noindent \bf Opening of schools:} In the base case, 
we assume that the schools open on July 1, 2021.

\bigskip

\begin{figure}
      \centering
     \includegraphics[width=\linewidth]{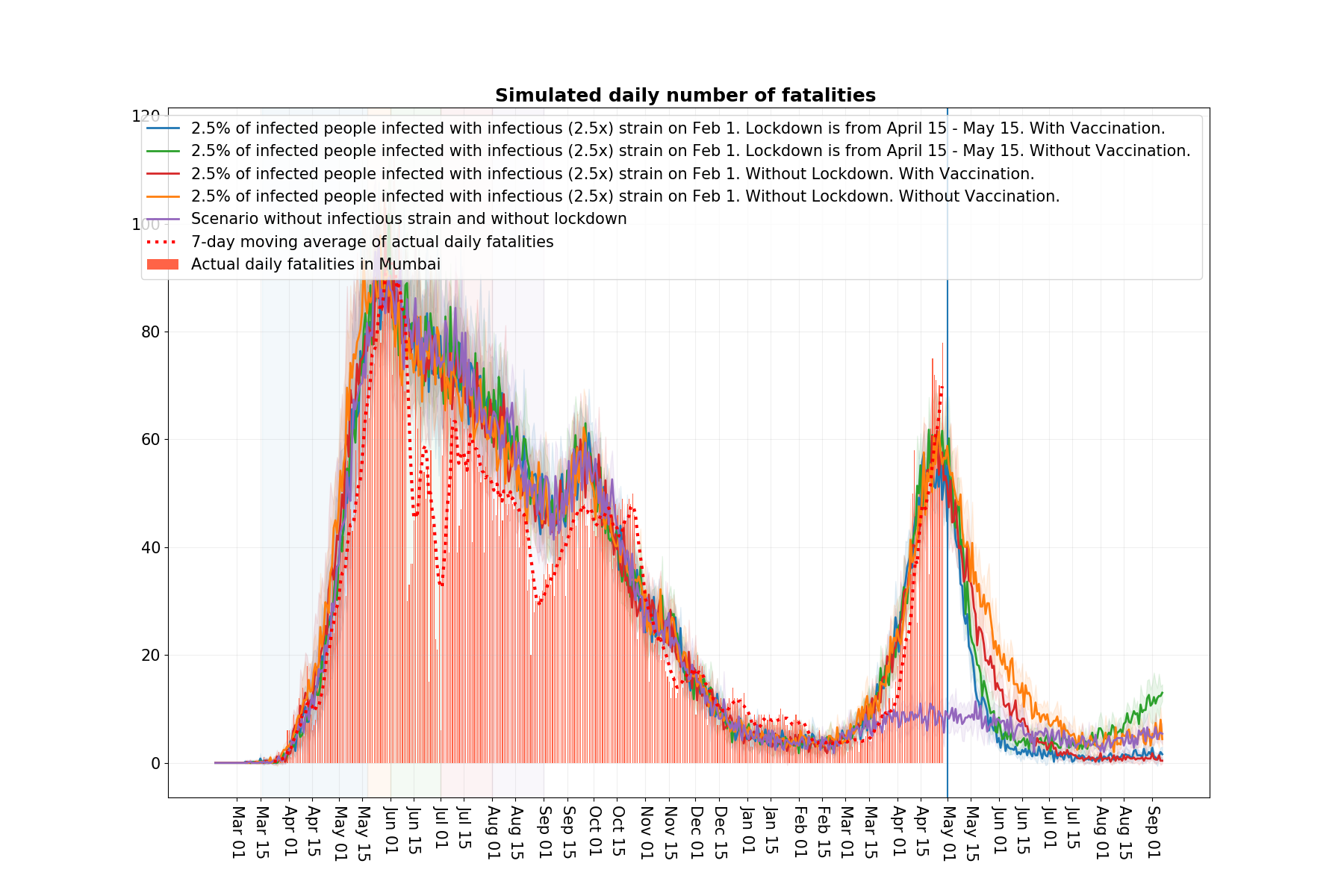}
      \caption{\small 
      Blue curve: Scenario with 2.5\% of infected people infected with infectious strain on Feb 1 (2.5 times more infectious). With lockdown from Apr 15 - May 15. And with vaccination effectiveness 0.75  (40\% above 60 yrs in Apr, 15 lakhs above 45 yrs in May, 20 lakh above 18 yrs each in June, July and Aug).  Orange curve: Scenario without lockdown and without vaccination. Green curve: Scenario with lockdown and without vaccination.  Red curve: Scenario without lockdown but with vaccination. Violet curve : Scenario without infectious strain and without lockdown. 
 } \label{daily_deaths_gen}
  \end{figure}

\begin{figure}
      \centering
     \includegraphics[width=\linewidth]{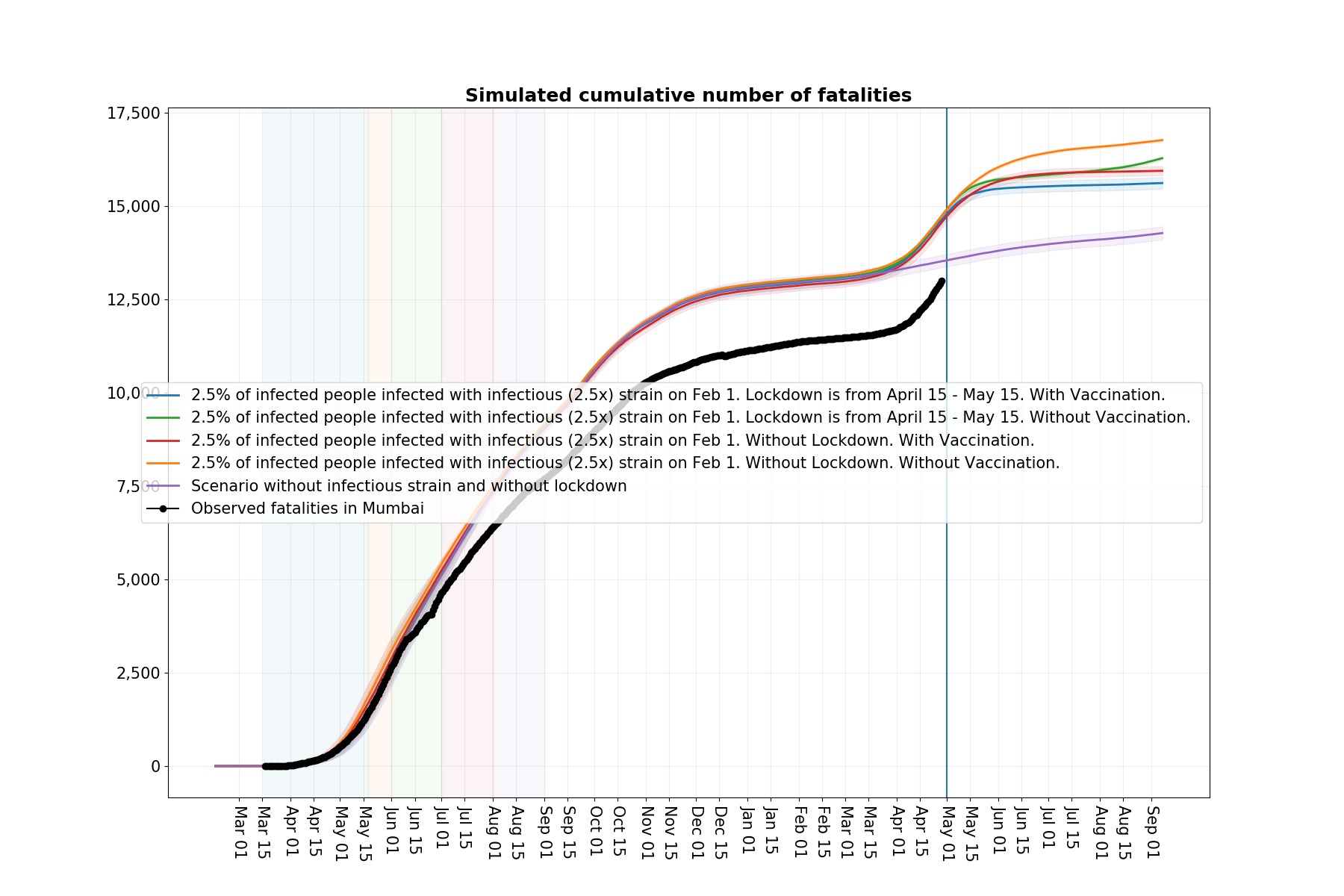}
      \caption{\small 
      Blue curve: Scenario with 2.5\% of infected people infected with infectious strain on Feb 1 (2.5 times more infectious). With lockdown from Apr 15 - May 15. And with vaccination effectiveness 0.75  (40\% above 60 yrs in Apr, 15 lakhs above 45 yrs in May, 20 lakh above 18 yrs each in June, July and Aug).  Orange curve: Scenario without lockdown and without vaccination. Green curve: Scenario with lockdown and without vaccination.  Red curve: Scenario without lockdown but with vaccination. Violet curve : Scenario without infectious strain and without lockdown. 
       } \label{fatalities_gen}
  \end{figure}

\begin{figure}
      \centering
     \includegraphics[width=\linewidth]{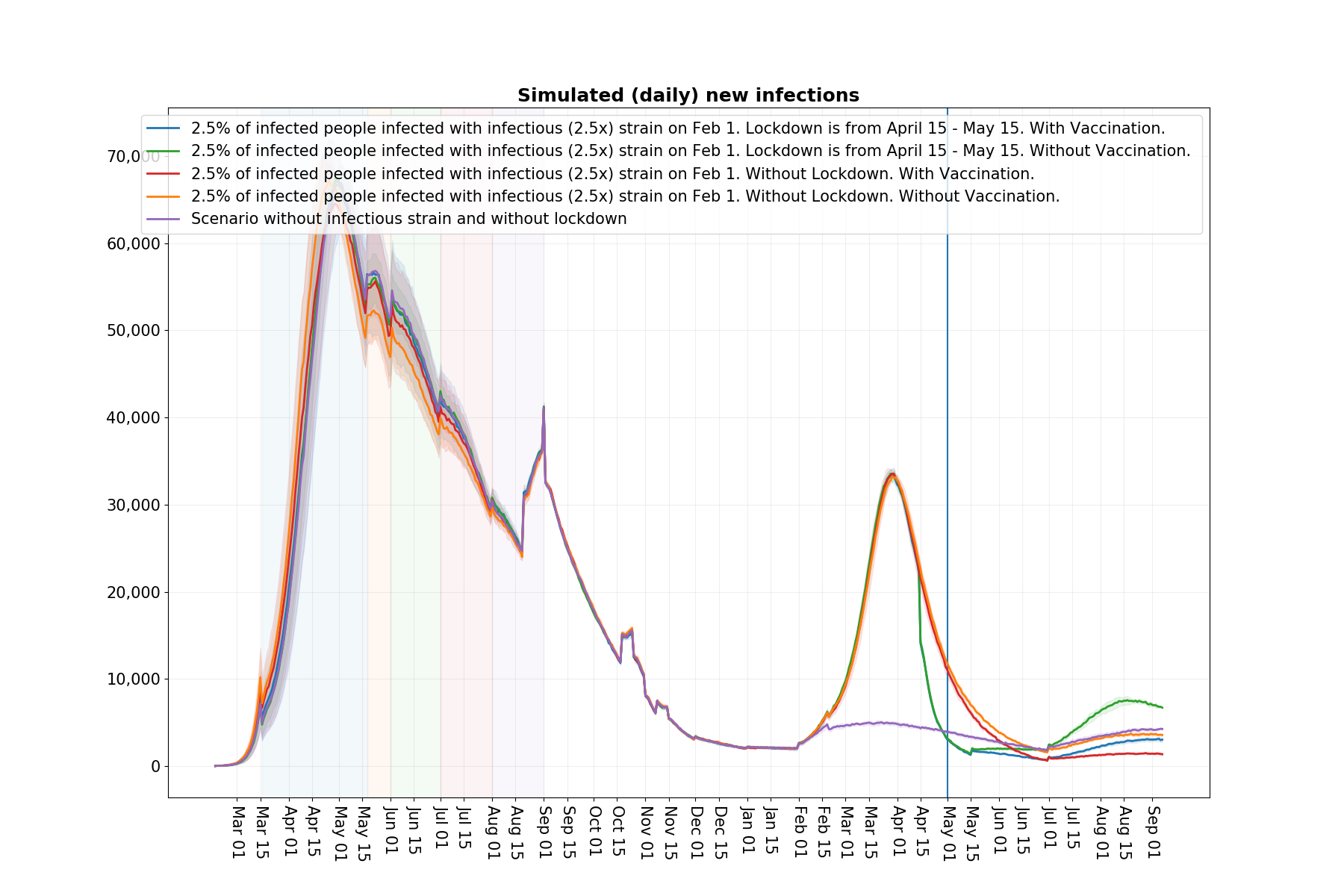}
      \caption{\small 
      Blue curve: Scenario with 2.5\% of infected people infected with infectious strain on Feb 1 (2.5 times more infectious). With lockdown from Apr 15 - May 15. And with vaccination effectiveness 0.75  (40\% above 60 yrs in Apr, 15 lakhs above 45 yrs in May, 20 lakh above 18 yrs each in June, July and Aug).  Orange curve: Scenario without lockdown and without vaccination. Green curve: Scenario with lockdown and without vaccination.  Red curve: Scenario without lockdown but with vaccination. Violet curve : Scenario without infectious strain and without lockdown. } \label{daily_infections_gen}
  \end{figure}

\begin{figure}
      \centering
     \includegraphics[width=\linewidth]{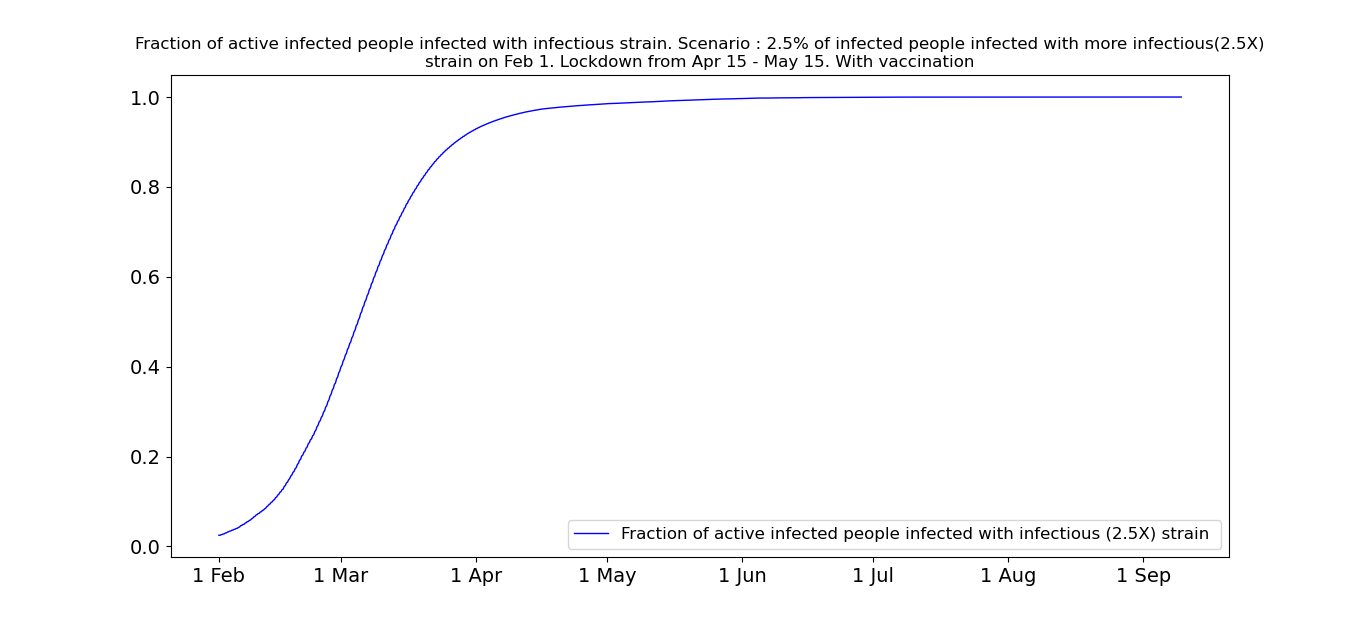}
      \caption{\small \text{Fraction amongst the infected with the highly infectious strain in the base case.}
      } \label{ratio_new_strain}
  \end{figure}

See Figures \ref{daily_deaths_gen}, \ref{fatalities_gen}, \ref{daily_infections_gen} and \ref{ratio_new_strain} for the results associated with the base case. 
Recall that the base scenario includes both the benefits 
of the  vaccination effort and the ongoing lockdown. 
In the graphs, the base scenario is compared with the following 4 scenarios:
\begin{itemize}
    \item 
    Scenario without the infectious strain and without the lockdown. This is the contra-factual
    setting where the mild wave is caused in Mumbai only due to 
    the opening of the economy on February 1, and increased laxity in population before that.
    \item
    Scenario without vaccination and without lockdown. This helps us estimate the 
    combined  benefits of the lockdown and the vaccination effort.
    \item
    Scenario without lockdown and with vaccination.
    This helps  isolate the benefits of the lockdown.
    \item
    Scenario with lockdown and without vaccination. This helps isolate the benefits 
    of the vaccination  effort.
\end{itemize}

The following configurations match those in \cite{October_report_2020} and  are common to all the figures hereafter
unless the perturbation is specified:  Workplace opening schedule is
     5\% attendance from May 18 to May 31st, 2020, 15\% attendance in June, 25\% in July, 33\% in August, 50\% from September 2020 to January, 2021 and 65\% from February, 2021. Schools open from 
     July 1, 2021. 
     
     Compliance levels are 0.6 (NS), 0.4 (S) before December and outside festivals. These change to (0.5,0.3) in 
     December, (0.4,0.2) in January, (0.2,0.1) in February 1-18 and (0.4, 0.2)  from Feb 19 to April 14. Train scaling is set to 0.40.
     
     \bigskip

 {\noindent \bf \large Some observations suggested from the base case:}
 Figure \ref{daily_deaths_gen} suggests that the lockdown substantially speeds
 up the return to normalcy.  Under the  blue curve one
 sees that daily fatalities return to Jan-Feb levels by June 1 due
 to the lockdown and vaccinations. The red curve corresponds to no lockdown and vaccinations and in this scenario
 the normalcy  is delayed by a month.
 Importantly, vaccines make this normalcy lasting. Else, as the green curve shows, the fatality numbers
 would have started to increase from mid July onwards. Comparing the violet
 and the blue curve in Figure \ref{fatalities_gen} suggests 
 that the more infectious strain may have cost the city from 1,500 to 2,500 extra fatalities by September 2021. 
 Figure \ref{daily_infections_gen} maps the projected daily infections in different scenarios.
 It suggests that even under the blue curve base case, we may see a few thousand new infections
 each day in September. Various sero-surveys suggest in Indian metropolitan cities roughly that 15-30 infections result in 
 one reported case, so that we may still see a few hundred cases daily in September. 
 Figure \ref{ratio_new_strain} illustrates  the speed with which a highly infective
 variant can come to dominate once the economy starts to open.

\clearpage

\subsection{Perturbed scenarios around the base case}

The scenarios around the base case that we consider include (these are summarized in Table \ref{Scenarios_considered_around_base}):

\begin{figure}
  \begin{center}
    \small
    \begin{tabular}{|c|c|}
      \hline
      {\bf Scenario tested} & {\bf Cases considered} \\
      \hline
      Economy opening on Feb 1 & 65\%, 75\% \\
      Compliance & Base Case, weaker compliance (2.5x and 2x infectious strain)\\
	  Reinfection & 0\%, 5\%, 10\%, 2.5\% each month (Feb-Jul)\\
	  Reinfection without infectious strain & 20\%, 5\% each month (Feb-Jul)\\
      Initial fract. of infected people with new strain & 1.5\%, 2.5\%, 5\%, 10\%\\
	  Infectiousness of new strain & 1.5x, 2x, 2.5x, 3x \\
	  Trains scale & 0.19, 0.40, 0.80\\
	  Vaccine effectiveness & 75\%, 55\%, No vaccination\\
	  Virulence of New Strain & Same as original strain, 1.3 times the original strain\\
	  Schools & Opening from July 1, Remain closed \\
      \hline
    \end{tabular}
   \end{center}
  \caption{Different scenarios considered around the base case.  }
  \label{Scenarios_considered_around_base}
\end{figure}

\begin{enumerate}
\item
{\bf Economy opening up:}  Figure \ref{daily_deaths_eco} shows the fatality curve when the 
city opened up to level 75\% on February 1 (red curve) instead of the base case (blue curve) where it opened
at a level 65\%. This leads to an increase in fatality levels from around March 1 onwards
and somewhat earlier peak time for the fatalities. 
This illustrates the broader phenomena that increasing or reducing
economic activity would lead to a shift in the fatality curve that does 
not help align the base curve with the steepness of the observed curve. 
A variable shift, where we first reduce the economic activity and later increase it may help in a better 
curve fit, but that  does not match with our experience of the city situation. 

  \begin{figure}
      \centering
     \includegraphics[width=\linewidth]{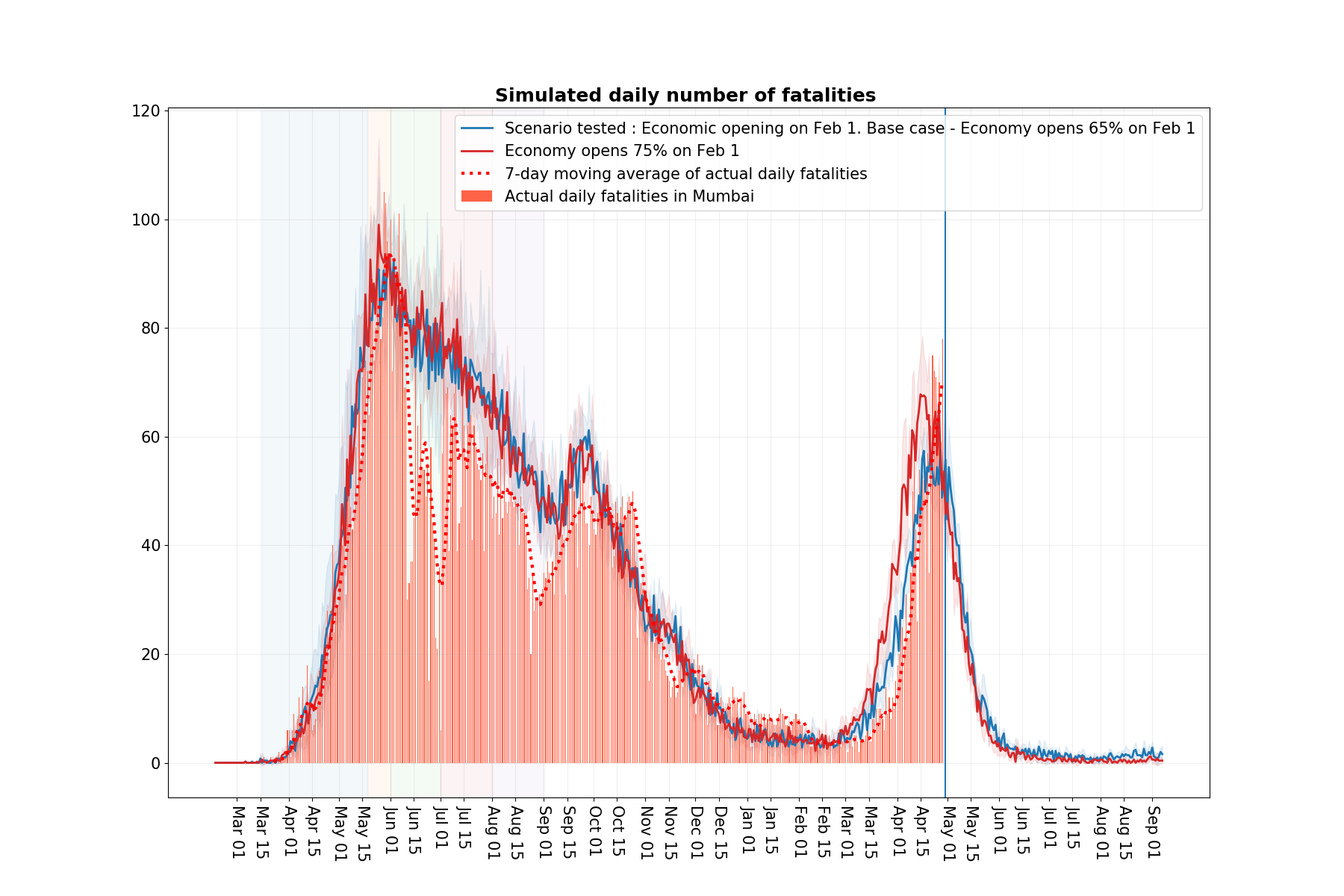}
      \caption{\textbf{Scenario: Opening up of the city.}
      Blue curve: 65\% economy opening on Feb 1. Red curve: 75\% economy opening on Feb 1.
  } \label{daily_deaths_eco}
  \end{figure}

\item
{\bf Compliance:}
In Figure \ref{daily_deaths_comp}, we show
a lower compliance scenario (red curve) and compare it to the base scenario (blue
curve). Again, the red curve leads to an increase from March 1 onwards in the fatality curve.
Its not clear how to reduce or increase compliance in a realistic manner that would achieve the steepness
of the observed fatality data. 

In Figure \ref{daily_deaths_comp_inf} we consider a more promising scenario
where the compliance is lowered as before, however the variant virus infectiousness is
reduced to 2 times from 2.5 times. This matches the data equally well except at the peak.
Further lowering compliance levels however may be an unrealistic match to reality.
Nonetheless, this example illustrates that infectiousness of the new variant
of order 2 is also consistent with the data. Matching any curve too closely
leads to over fitting and is not desirable. Further, peak fatality data around mid to end April may be high 
due to avoidable fatalities as Mumbai medical infrastructure was 
quite stretched around mid-April, so we do not look for a close match there.

  \begin{figure}
      \centering
     \includegraphics[width=\linewidth]{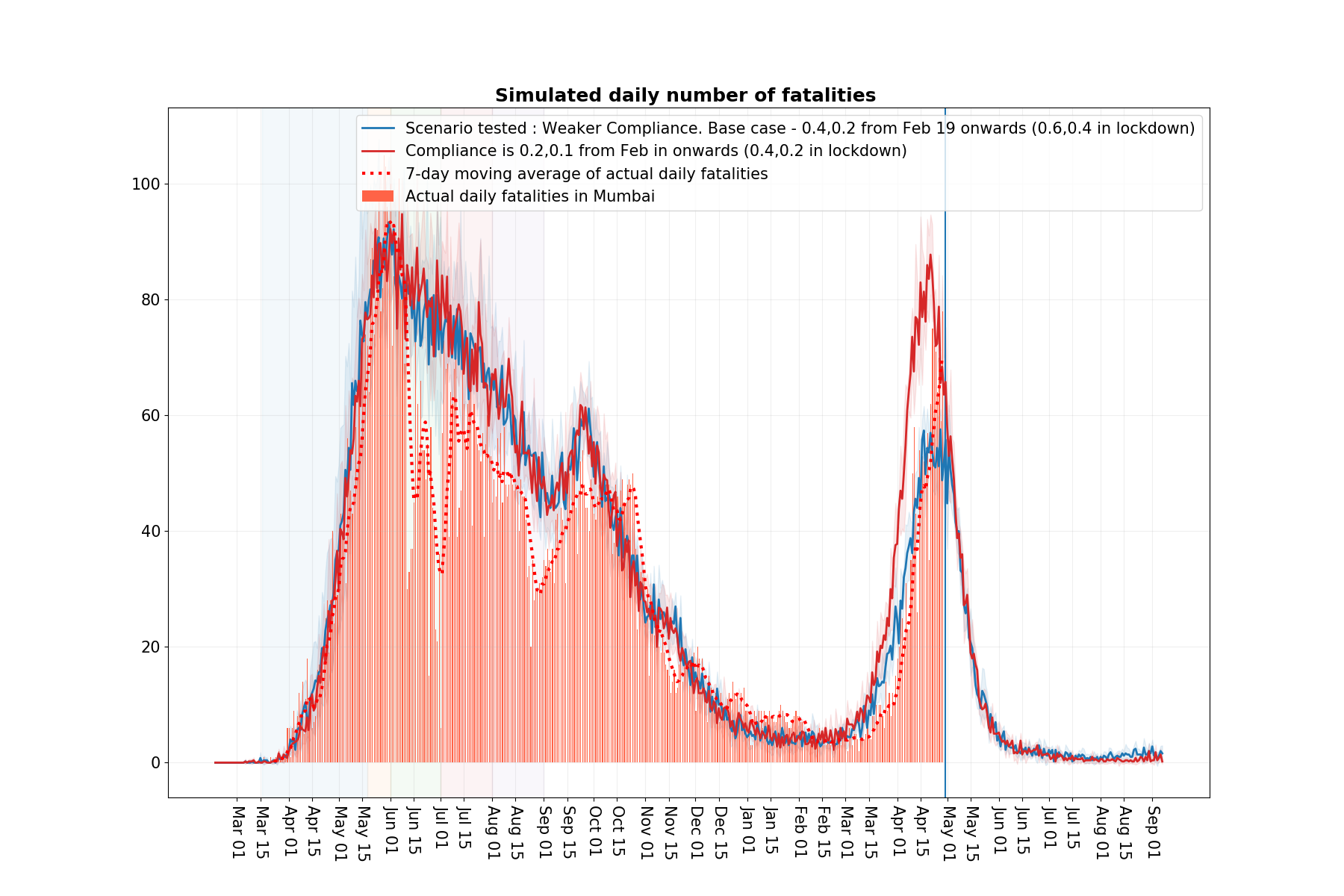}
      \caption{\textbf{Scenario: Variable compliance.}
      Blue curve: Compliance in this scenario is 0.4 in non slums and 0.2 in slums Feb 19 onward and 0.6,0.4 in the lockdown. Red curve:  0.2, 0.1 from Feb 19 to Apr. 14.  In the lockdown (Apr 15 - May 15) it is set to 0.4, 0.2 . } \label{daily_deaths_comp}
  \end{figure}

  \begin{figure}
      \centering
     \includegraphics[width=\linewidth]{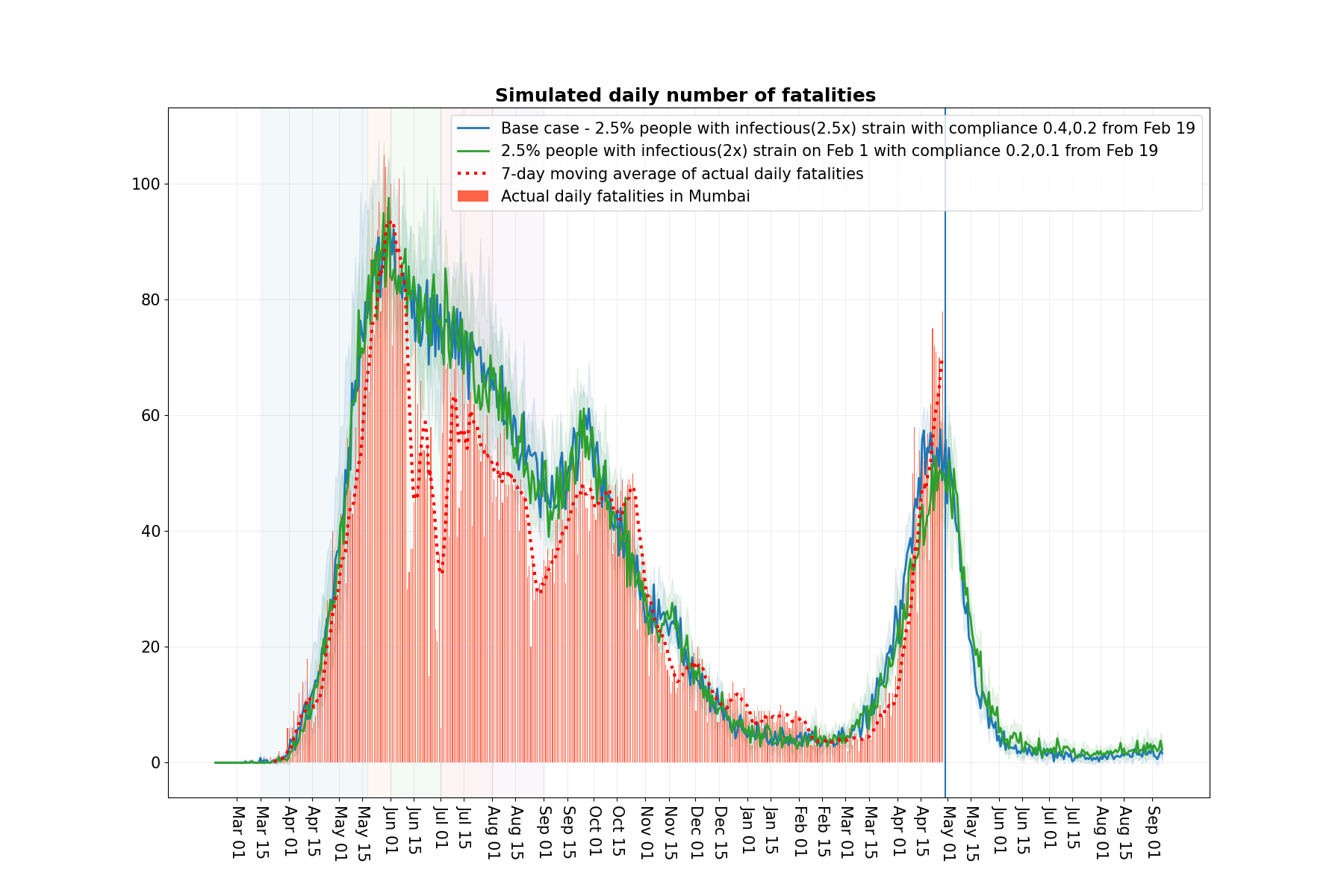}
      \caption{\textbf{Scenario: Lower compliance and (2x) infectious strain.}
      Blue Curve: Compliance in this scenario is 0.4 in non slums and 0.2 in slums Feb 19 to Aprl 14 and 0.6, 0.4 in the lockdown. Infectiousness of new strain is 2.5 times the original strain. Green curve: Compliance  0.2, 0.1 from Feb 19 to April 14.
        In the lockdown (Apr 15 - May 15) it is set to 0.4, 0.2. Infectiousness of new strain is 2 times the original strain. } \label{daily_deaths_comp_inf}
  \end{figure}

\item
{\bf Reinfection:} In Figure \ref{daily_deaths_reinf}
we consider the cases where reinfection is 5\% (green curve) on February 1.
Technically, this means that we convert randomly chosen 5\% of the recovered population
on February 1, and treat it thereafter as  susceptible. Red curve
captures the case where the reinfection is set to 10\% on
February 1. As Figure \ref{daily_deaths_reinf} shows, increase in reinfection as specified
simply leads to higher fatality numbers that more-or-less increase
linearly from March 1 with a high slope.
In the orange curve we consider a case 
where reinfections are introduced gradually
at 2.5\% each month from February 1 to July 1.
The curve again increases very steeply. 
While these are some ad-hoc numbers, its clear
that reinfections would need to increase
in a very specific manner and combine with appropriate compliance  and variant evolution 
to result in a curve that
matches the observed fatality curve.

In Figure \ref{daily_deaths_only_reinf}
we assume that the variant does not exist
and play only with the reinfection scenarios.
The green curve corresponds to 20\%
reinfection amount on February 1.
The red curve considers the case where reinfections are introduced gradually
at 5\% each month from February  1 to July 1. These curves 
belabour the point that the fatality
data can be explained
using only reinfections
in a  very specific manner, involving much large number of infections 
closer to late March compared to those in early March. A pattern of
that sort does not appear natural and does not suggest  itself
from the anecdotal reports of reinfected cases,
and thus appears unlikely.

Another way to interpret  converting the recovered population to susceptible on  February 1  
in our model is that it  directionally accounts for the possible underestimation of the  
 susceptible population  on February 1 by the model.  The fact that even in these settings, the fatality numbers
come down to earlier January and February levels on June 1 is reassuring
for our projections.

  \begin{figure}
      \centering
     \includegraphics[width=\linewidth]{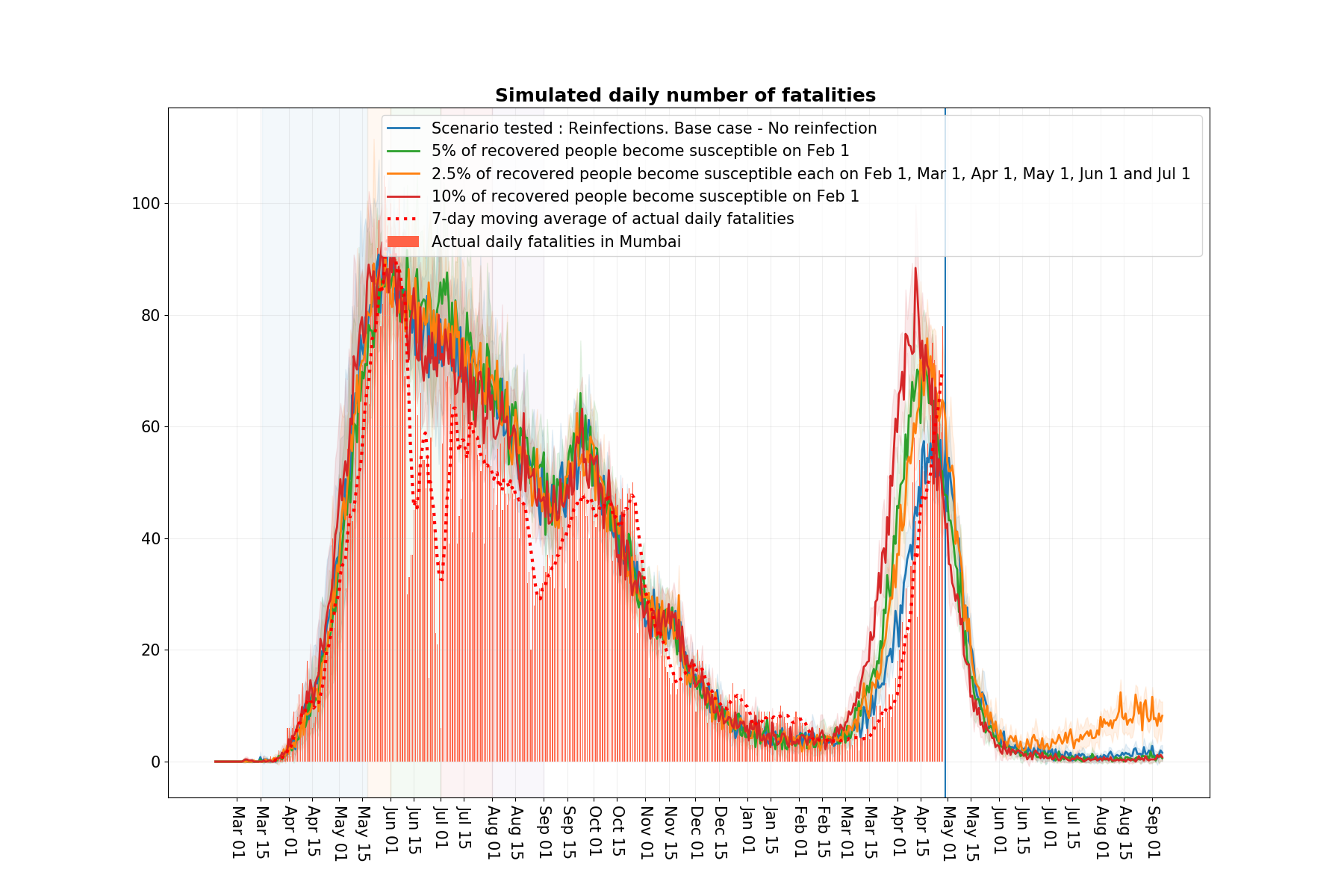}
      \caption{\textbf{Scenario: Reinfections.}
      Blue curve: No reinfections. Red curve: 10\% of recovered people eligible for reinfection on Feb 1. Green curve: 5\%. Orange curve: 2.5\% each on Feb 1, Mar 1, Apr 1, May 1, Jun 1, Jul 1.} \label{daily_deaths_reinf}
  \end{figure}

  \begin{figure}
      \centering
     \includegraphics[width=\linewidth]{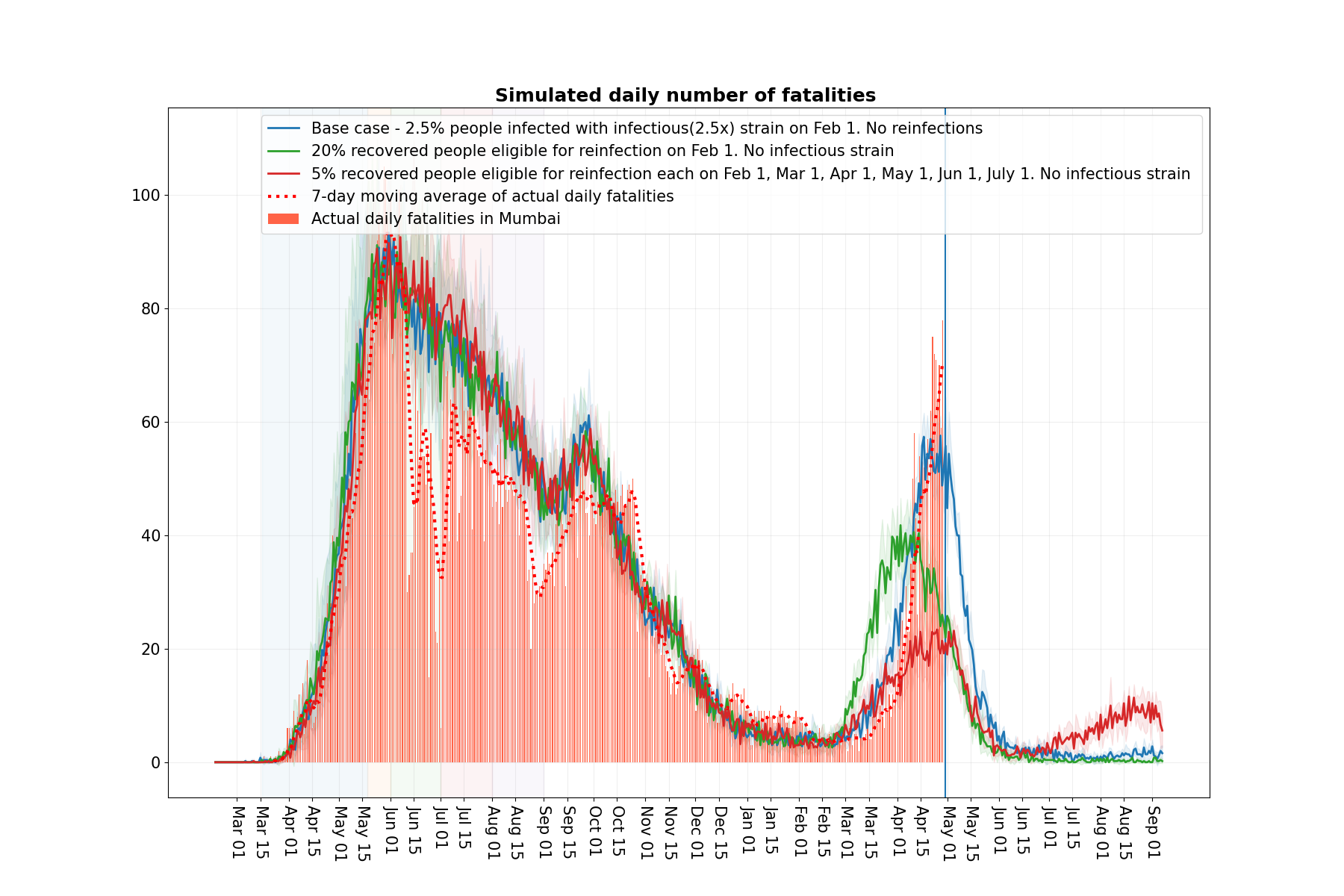}
      \caption{\textbf{Scenario: Reinfections. The more infectious strain is absent.}
     } \label{daily_deaths_only_reinf}
  \end{figure}

\item
{\bf Fraction of Infected people infected with the new strain on Feb 1 : }
%To get a first order idea of the impact of reinfections,
%we consider settings  where 5\% 
%of all  patients who had recovered by February 1 are now susceptible to the virus
%On Feb 1 this 5\% of recovered correspond to x \% of still susceptible (DM).
%Otherwise, our model assumes that the recovered 
%%can no longer be infected. 
%Alternative interpretation of this 5\% population is that 
%our model had over estimated the number protected 
%through a prior infection, and we check the sensitivity %of the model output to this error.

In Figure \ref{daily_deaths_people},
the red curve corresponds to the case
where the new strain is assigned 10\% of the infected on Feb 1.
The orange curve corresponds to 5\% , the green curve corresponds to 1\%,
and blue curve with 2.5\% denotes the base case. 
These curves suggest that values close to but smaller
than 2.5\% on Feb. 1 perhaps with slightly higher infectivity 
may match the observed data a little better than the base case.
 However, our broad conclusions are unchanged by this.

  \begin{figure}
      \centering
     \includegraphics[width=\linewidth]{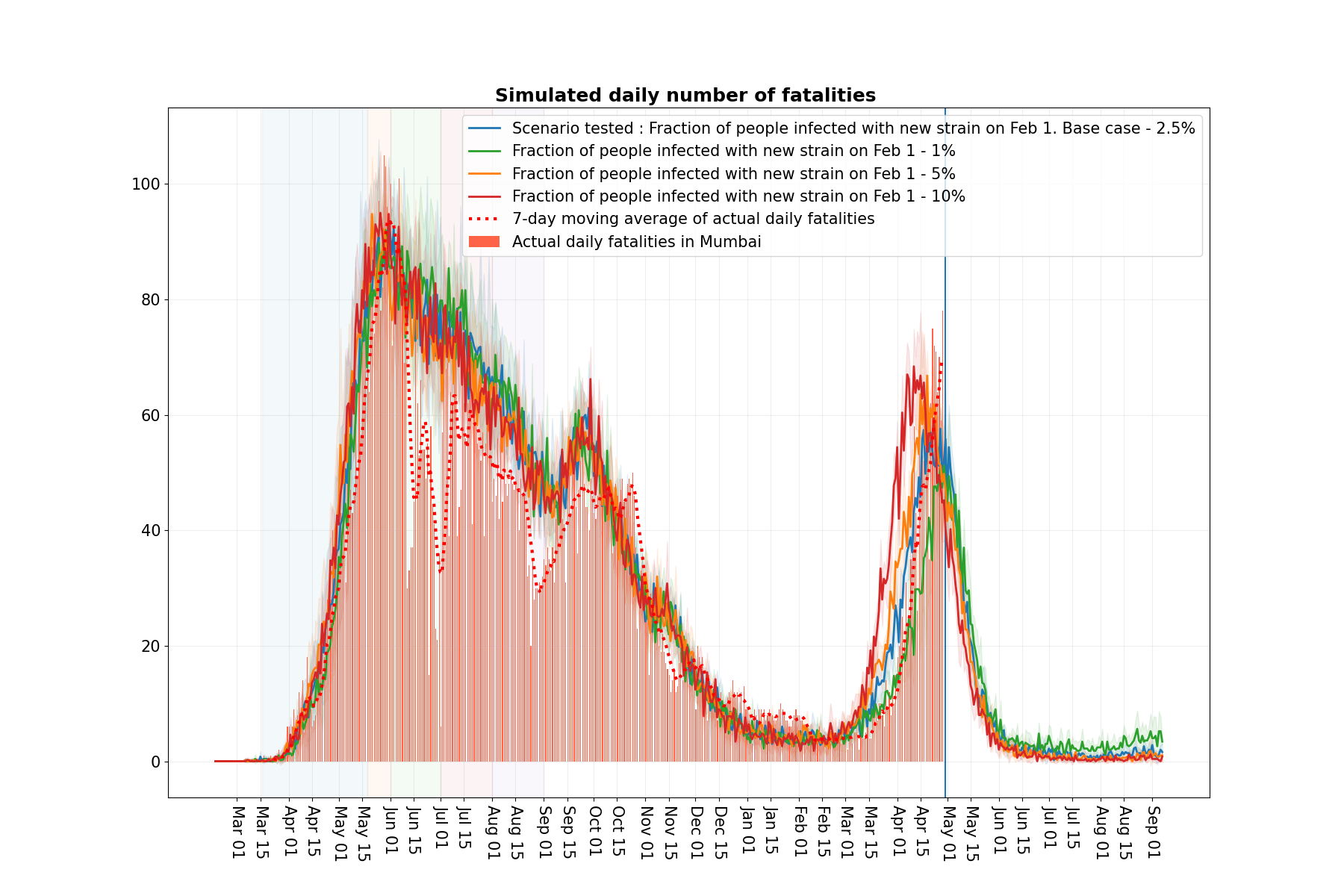}
      \caption{\textbf{Scenario: Fraction of infected people infected with new strain.}
      Blue curve: 2.5\% of infected people infected with infectious strain on Feb 1. Red curve: 10\% . Orange curve: 5\% . Green curve: 1\% . } \label{daily_deaths_people}
  \end{figure}

 \item
 {\bf Infectiousness of new strain : } In the base scenario we considered that the new strain was 2.5 times 
 more infectious than the original strain.  In Figure \ref{daily_deaths_inf}, we compare this base scenario with the scenarios where new strain is 1.5 times (green curve), 
 2 times (orange curve) and 3 times (red curve) more infectious than the original strain.
 The curves reaffirm that  2.5 times infectiousness (blue curve) is a good fit compared to the neighbouring values.

\item 
{\bf Trains:}
To
capture the varying levels in infection spread through trains,
in Figure \ref{daily_deaths_trains},
we consider the scenarios of low $\beta_T = 0.19\beta_H$ (green curve)
and high $0.8\beta_H$ (red curve). Both the curves
provide a similar fit compared to the base case.
The high infection  rate red curve matches the steepness of observed fatalities better,
however, it results in much higher fatalities in July and August 2020 than were actually observed.
These curves generally support the idea 
that an infective variant grew quickly in February and March, and that trains may have played
 a role in the infection  spread and in its  growth pattern.

\item
{\bf Vaccination:}
To capture the scenario the vaccines may have lower effectiveness, as well to indicate the effect of 
the vaccination drive not being as extensive as in the base case,
in Figure \ref{daily_deaths_vac}, we consider the case where
vaccine efficacy is 55\% (green curve) and where it is zero (red curve). 
Recall that in the base case the schools open in July. Figure  \ref{daily_deaths_vac}
suggests that even a moderately successful vaccination drive will help keep
the fatality numbers low in August and September from the `third wave'
that may otherwise result from school opening.
Again, as we noted earlier, the decision to open the schools on July 1 or later
is best made closer to those times, when one has a better idea of the
infections resulting from the opening.

\item
{\bf Variant virulence:}
Figure \ref{daily_deaths_vir} shows the scenario
where the new strain virulence or fatality rate is set to 1.3 times that of the original strain (red curve). 
This appears to better match the observed fatality data compared to the base case (blue curve)
around the peak values in late April (when the effect of the new strain is more pronounced
as it becomes dominant around late March). Again, since some of those high values may be due to avoidable deaths,
it is difficult to claim from the experiments that the new strain maybe more virulent. This is also not suggested 
by the current Mumbai SCFR graph.

{\noindent \bf Implementation: } Technically, due to lack of medical data, the increased virulence  is implemented in our model by  increasing the probability of an individual 
transitioning from symptomatic state to hospitalised state, from hospitalised to critical state,
from critical state to fatality, each 
by a factor of  cube root of 1.3.

\item
{\bf Schools:}
In Figure \ref{daily_deaths_school}, we compare the base case (blue curve) where the schools open from July 1 to
the case where they remain closed (red curve). While the fatalities show a very minor difference between the two
curves,
Figure \ref{daily_new_infections_school}, shows that the daily infections under the base case may be of the order
of 2,000 to 4,000. Again, given that typically 15-30 infections lead to a single reported case, this suggests that opening of schools may lead to a few hundred reported cases each day.
Since most of the vulnerable population would be vaccinated by then, assuming that vaccines 
are effective, this would translate to very few fatalities daily.

  \begin{figure}
      \centering
     \includegraphics[width=\linewidth]{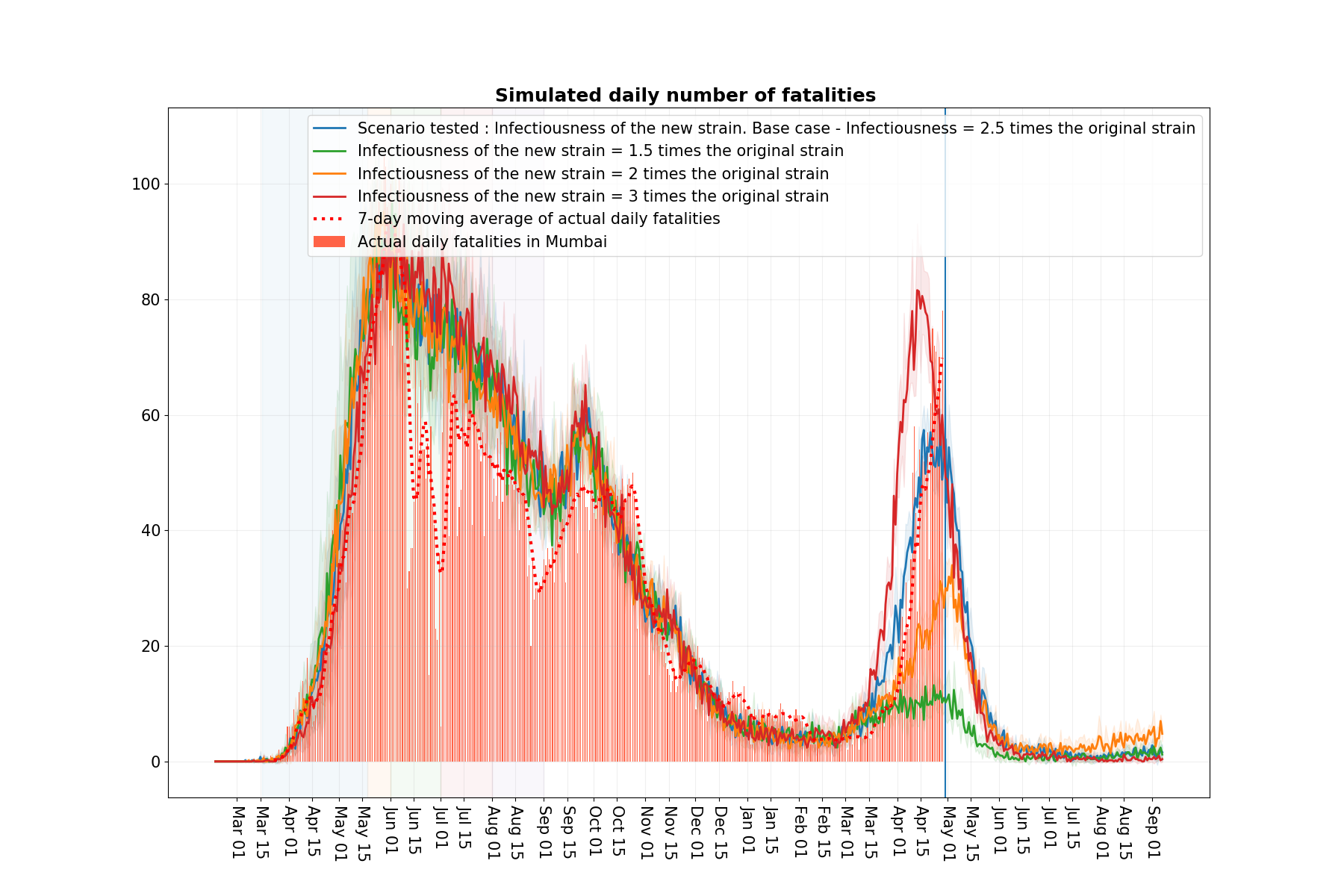}
      \caption{\textbf{Scenario: Infectiousness of new strain.}
      Blue curve: 2.5 times the original strain. Red curve: 3 times . Orange curve: 2 times . Green curve: 1.5 times . } \label{daily_deaths_inf}

     \includegraphics[width=\linewidth]{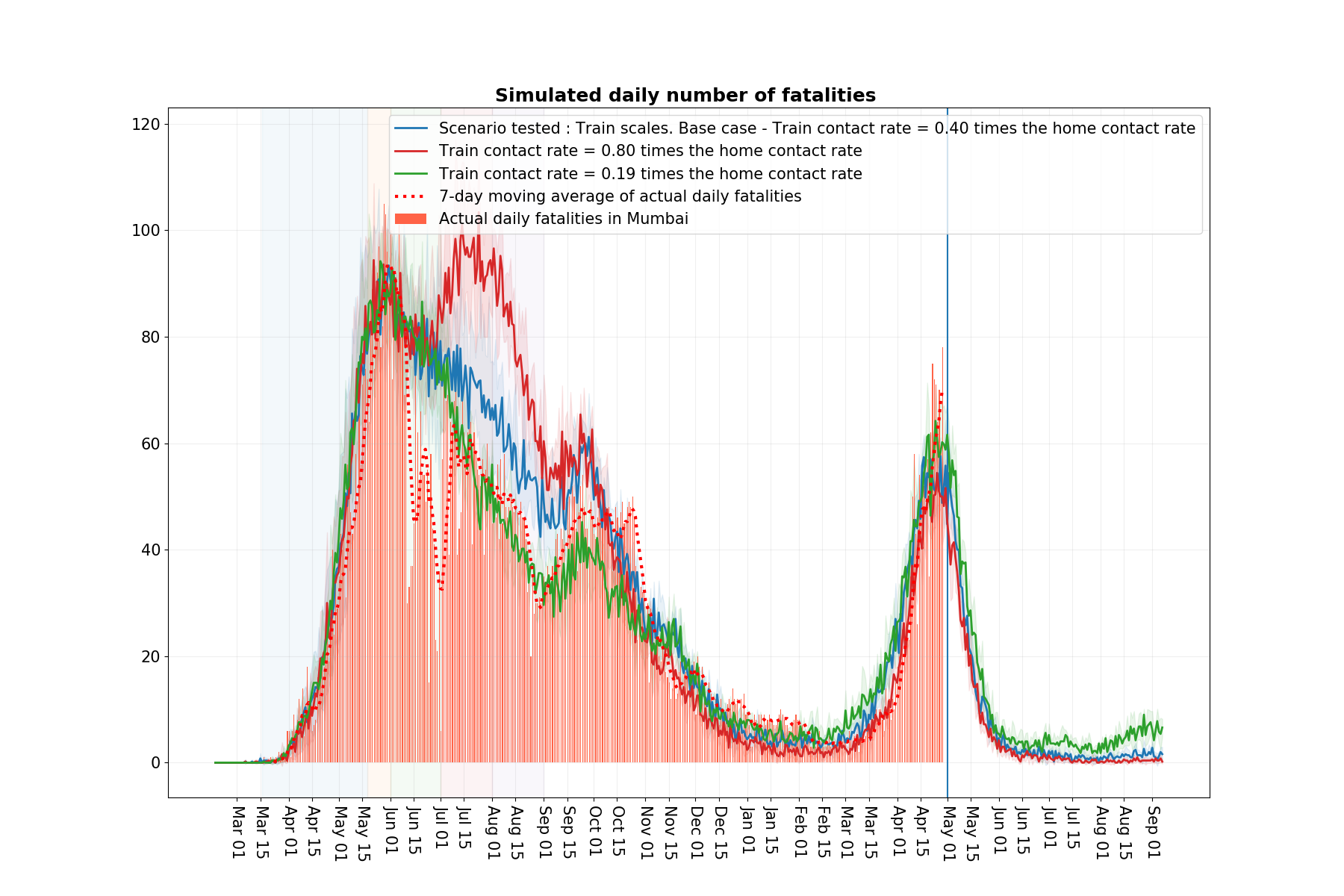}
      \caption{\textbf{Scenario: Train infectivity.}
      Blue curve (base): 0.40. Red curve (high): 0.80.  Green curve (low): 0.19. } \label{daily_deaths_trains}
  \end{figure}

  \begin{figure}
      \centering
     \includegraphics[width=\linewidth]{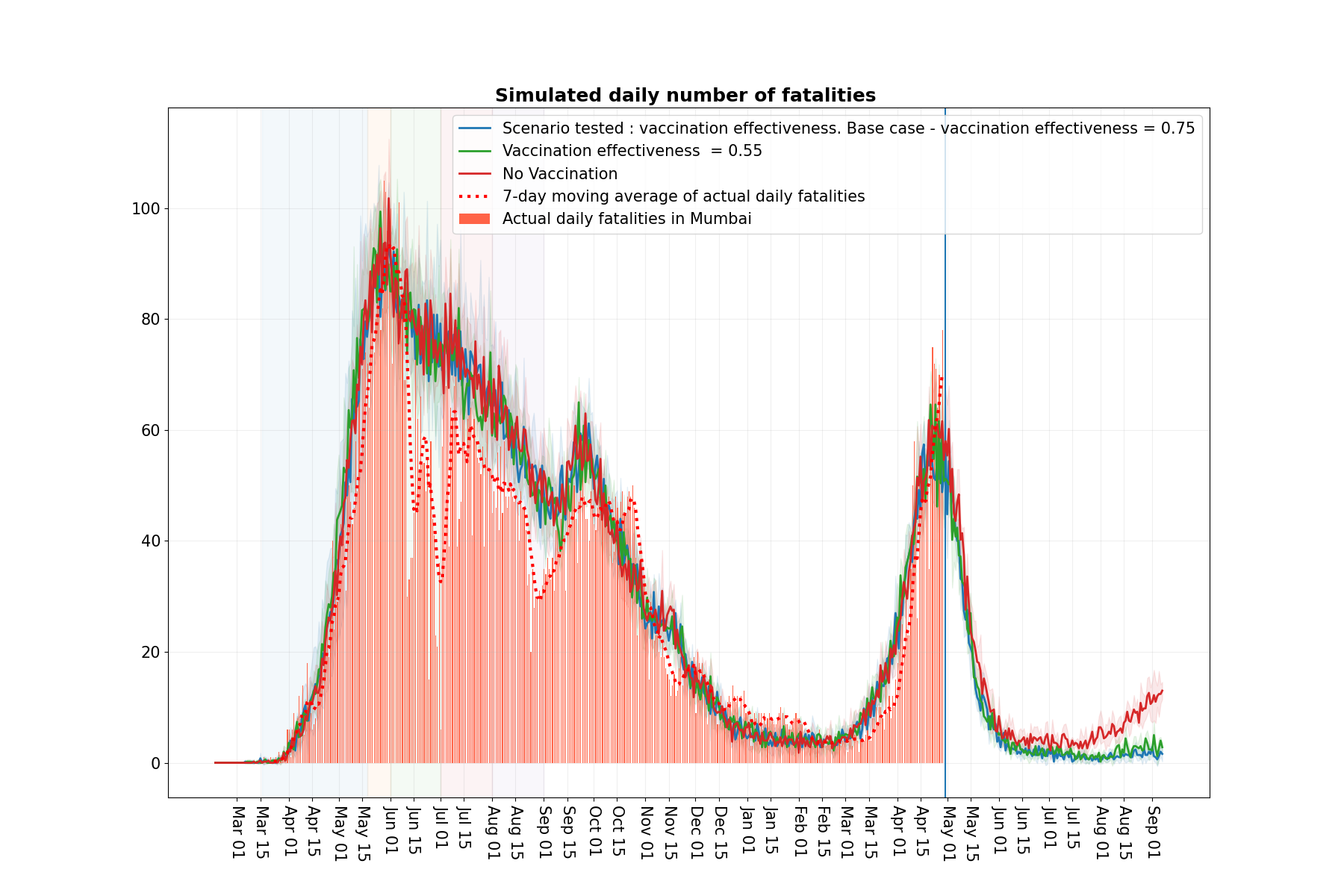}
      \caption{\textbf{Scenario: Vaccination effectiveness.}
      Blue curve: Vaccination effectiveness = 0.75. Green curve: 0.55 . Red curve: No vaccination. 
       } \label{daily_deaths_vac}

     \includegraphics[width=\linewidth]{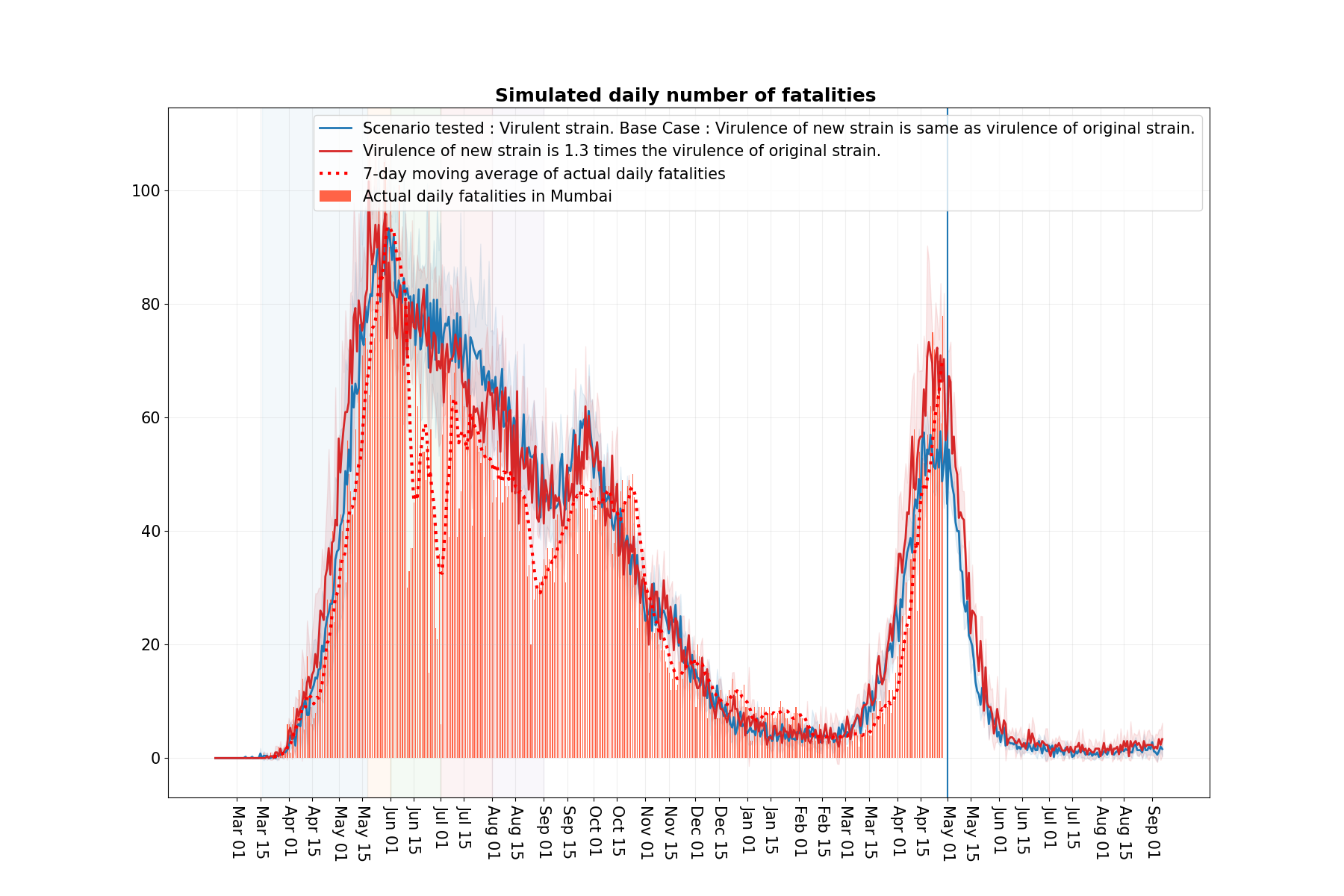}
      \caption{\textbf{Scenario: Variant virulence.}
      Blue curve: Virulence is same as original strain. Red curve: Virulence is 1.3 times the original strain. 
      } \label{daily_deaths_vir}
  \end{figure}

{
  \begin{figure}
      \centering
     \includegraphics[width=\linewidth]{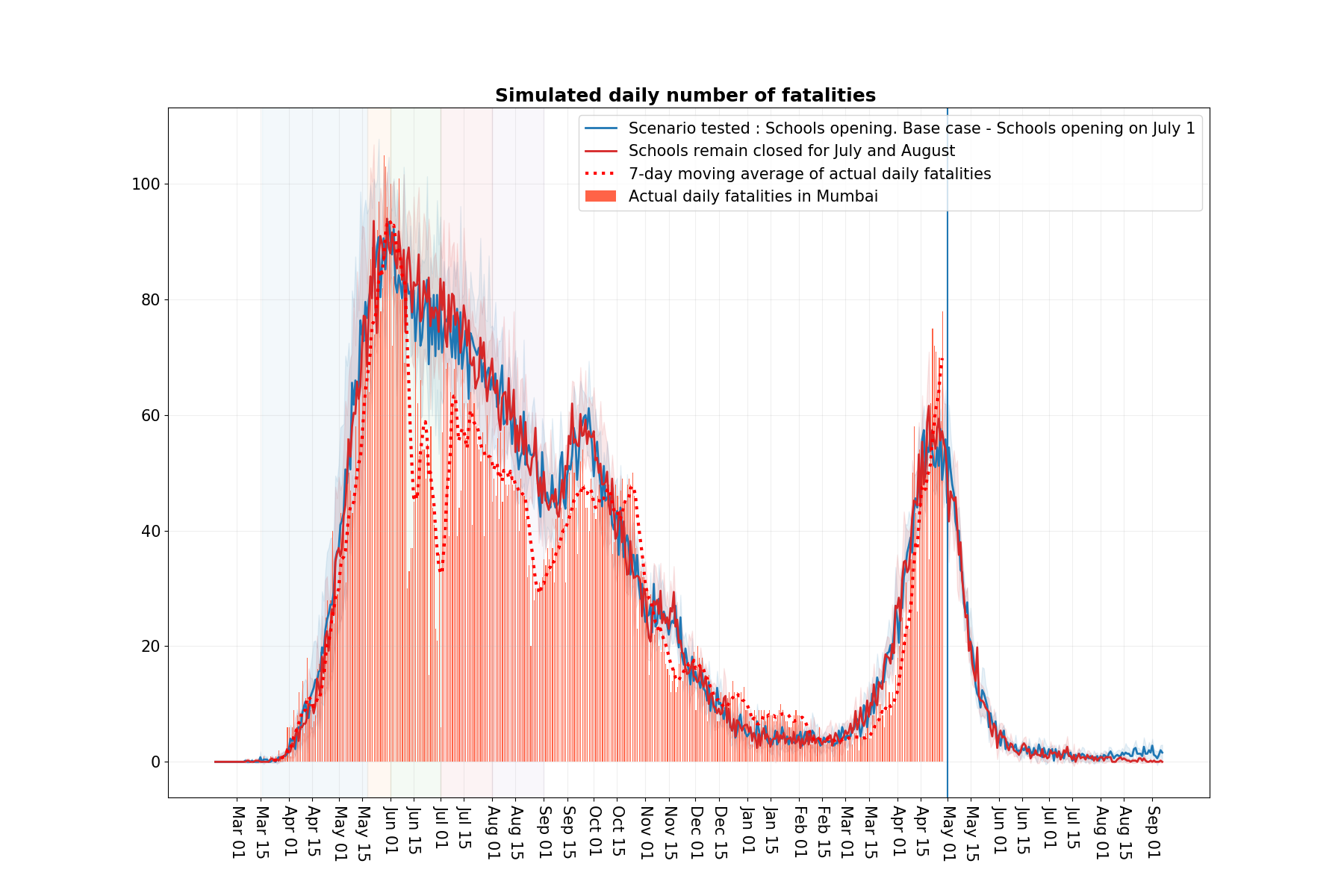}
      \caption{\textbf{Scenario: Schools opening.}
      Blue curve: Schools open from July 1. Red curve: Schools remain closed in July and August. 
       } \label{daily_deaths_school}

     \includegraphics[width=\linewidth]{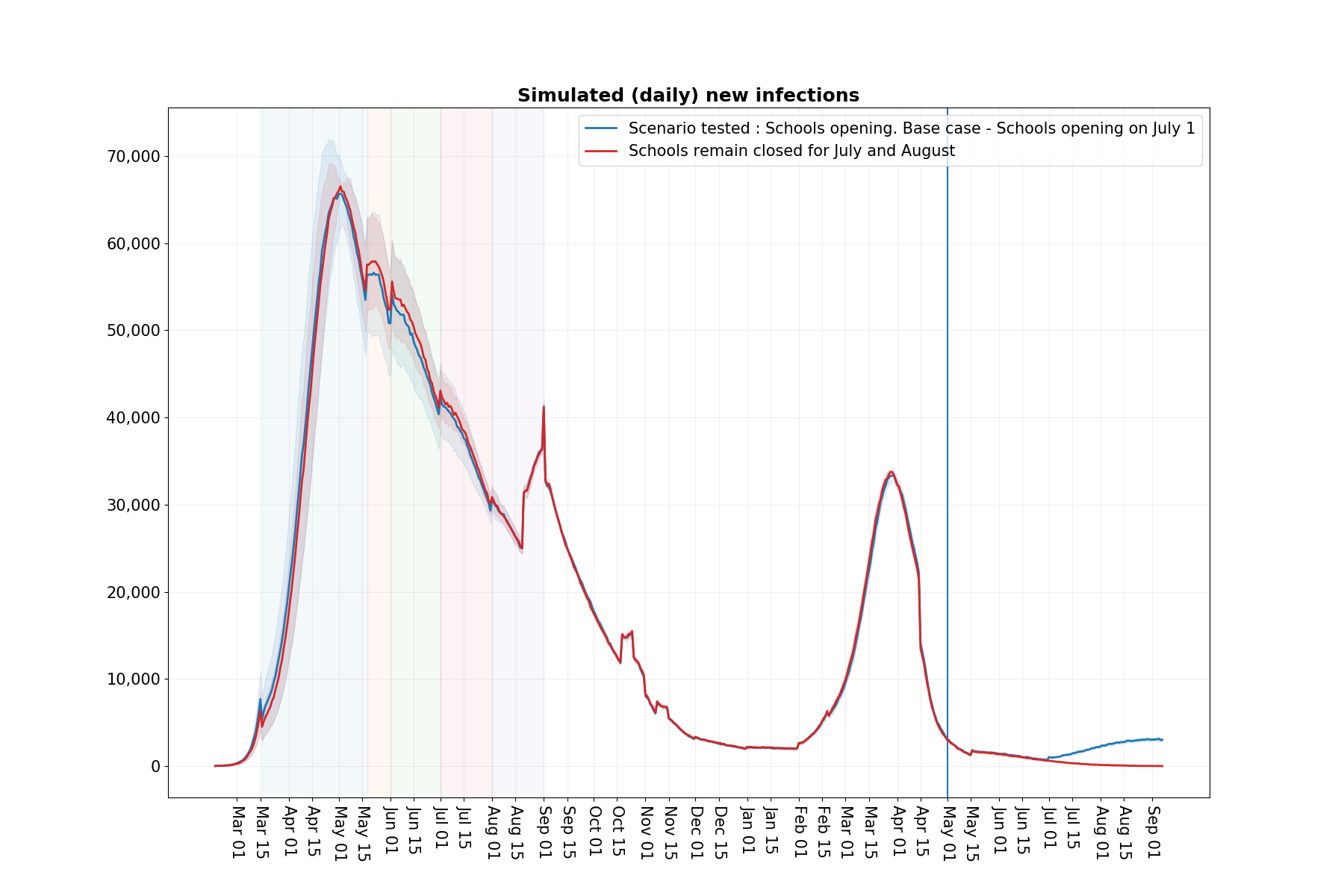}
      \caption{\textbf{Scenario: Schools opening.}
      Blue curve: Schools open from July 1. Red curve:  Schools remain closed in July and August. 
       } \label{daily_new_infections_school}
  \end{figure}
  }

\end{enumerate}

So were Mumbai trains the key reason for the severe second wave in Mumbai?
It appears that while they certainly contributed to the increasing infections in the city, if they had led to 
a dramatic increase, that would have shown up in the fatalities observed in March (given that fatalities
lag exposure by roughly a month). Since, fatalities had a
phase change around end of March, this suggests that trains may not have been the key reason. 
Opening up of the economy at any nearby time before or after February 
would likely have led to growth in variants (since its unlikely that a large proportion of population would
have been vaccinated any time soon), and that  is the suggested key reason for the severe second wave as per our computational experiments. Mumbai trains certainly played an important role in expediting their spread.

\section{The $R_0$ in our model and other technical enhancements}
 \label{section:R_0} 
 
 Recall that $R_0$ denotes the expected number of individuals a single randomly selected exposed person infects
 in a city where everyone else is susceptible.  It captures the infectivity of the virus.
 All else being equal, a higher $R_0$ implies a more infective virus.
   Below, in Figure \ref{R_0_a}, we report the $R_0$ for our model when it is fitted to
 the fatality data last year, {\em the base case}. See \cite{City_Simulator_IISc_TIFR_2020}
 for details of how our model was fit to data. We also report the $R_0$  from a variant
 that in our model is two times or two and half times more
 infective in terms of the transmission rates compared to the base case. 
  
 The $R_0$ for overall city corresponds
 to the case where  the exposed individual is randomly selected from across the whole city.
 We also report $R_0$ when the exposed individual is randomly selected from a non-slum area 
 as well as from a slum area.  It is higher in the latter case, because in a more dense setting,
 an individual is likely to interact with more people
 and infect more of them.
 
Since the current infections in Mumbai are largely in non-slums,
the increase in $R_0$ from non-slums is a better measure 
of the impact of more infective variants to the city.

Figure \ref{R_0_a} 
suggests roughly that in the non slum areas the $R_0$ has increased from around 2 to 2.5 for the base case  to over
3 under the variants with transmission rates increased by a factor from 2 to 2.5. 
 
 \begin{figure}[h]
  \begin{center}
    \small
    \begin{tabular}{|c|c|c|c|}
      \hline
      {Relative infectiousness w.r.t. base case} & {$R_0$ for non slums} & {$R_0$ for overall city} & {$R_0$ (inferred) for slums} \\
      \hline
      1 & 2.17 $\pm$ 0.108 & 4.02 $\pm$ 0.27 & 5.66\\
      \hline
      2 &  3.135 $\pm$ 0.15 & 5.56 $\pm$ 0.28 & 7.71\\
      \hline
	  2.5 &  3.55 $\pm$ 0.17 & 6.56 $\pm$ 0.31 & 9.24\\
      \hline
    \end{tabular}
   \end{center}
  \caption{$R_0$ values for non slum areas, overall city and slum areas. 
  The confidence intervals capture the 95\% statistical error from our simulation sampling.}
  \label{R_0_a}
\end{figure}

Below we recall our simulation dynamics. These help illustrate the $R_0$ 
estimation procedure  as well as the methodology to incorporate 
more infective variants in our simulation model.

\subsection{Simulation dynamics}

Recall that our simulation  model  works as follows (see \cite{City_Simulator_IISc_TIFR_2020}):
\begin{itemize}
\item
 At a well chosen start time for our simulation, a  fixed number of exposed individuals are seeded in the city
 where every one else is susceptible. 
 \item
 The simulation proceeds iteratively over time, incrementing it by $\Delta t$ at each time step.
In our simulation $\Delta t$ corresponds to 1/4 of  a day, or six hours.  At each time $t$, 
for every susceptible individual   $n$,
its infection rate  $\lambda_n(t)$  is the sum of infection rates coming from
all the infected individuals in
 his interaction spaces including home ($h$), workplace ($w$), school ($s$), community spaces ($c$)  and 
 transport ($T$). There are other categories in our model and  they may be similarly handled.
  Thus, if $\lambda_n^{n',a}(t)$ denotes the rate at which individual $n'$ in the interaction space 
 $a\in (h,w,s,c,T)$ infects individual $n$ at time $t$ (this would be zero if $n'$ is not infective
 or not interacting with $n$), we have
\[\lambda_n(t)= \sum_{n',a \in (h,w,s,c,T)}\lambda_n^{n',a}(t).\]
\item
Then at time $t+\Delta t$, each susceptible individual moves to the exposed state with probability $1-\exp{(-\lambda_n(t) \Delta t})$, independently of all other events. With the remaining probability it continues to be susceptible. 
\item 
 Individuals
 once exposed, follow  a disease progression probabilistic dynamics 
 as specified in \cite{City_Simulator_IISc_TIFR_2020}. Some exposed become asymptomatic or symptomatic
 and they may infect others in the coming time periods.
 Asymptomatics recover after a short duration. Symptomatics may either recover 
 or a small  age dependent  fraction may be  hospitalised. 
 Hospitalised may recover
 or a small age dependent fraction may become critical. Critical cases may recover or 
 a small age dependent fraction may pass away.
\item
  Time is then incremented to $t+\Delta t$ and the condition of individuals
  are  updated.  The simulation iteratively continues till some large specified terminal time.
\end{itemize}

Above, if a susceptible individual becomes infected at time $t$, there remains an issue of identifying
the individual who  infected
this individual. In our algorithm, this {\bf blame} is assigned uniquely to one individual
$n'$. And this assignment happens  with probability 
\[
\sum_{a \in (h,w,s,c,T)}\lambda_n^{n',a}(t)/\lambda_n(t).
\]

This appears to be a fair blame allocation that can be shown to  
probabilistically asymptotically valid  as $\Delta t \rightarrow 0$.

\bigskip

{\noindent \textbf{Estimating $R_0$:} }
At day zero, a randomly selected  individual is marked as exposed to the disease, 
while all others are marked as susceptible. 
The selected individual follows the disease progression dynamics. 
When in an infective state, he may infect others.
 We count the total number of individuals infected by the selected individual until 
 he is no longer infective. The above specified allocation rule helps
 in arriving at this number uniquely. 
 This is a random quantity. This experiment is repeated independently
 many times to arrive at an estimator
 for $R_0$.
 
{\noindent \textbf{Modelling the more infectious strain:} }  
\textbf{} 
At a well chosen time, in our case February 1, 2021, a randomly selected fraction of 
the
infectious are selected and marked as having a variant. Their infection rates to other individuals
are accordingly bumped up by the infective factor. So if the virus is two times more infective, the corresponding rate
increases by a factor 2. 
Our simulation algorithm then proceeds over time 
with small changes. For each infected individual we keep track of whether it is infected 
with an old or a new strain. Further, suppose that 
at  time $t$ when a susceptible individual $n$
gets infected, we need to decide whether the infection is from
the original strain or from the new one. If $\lambda_n^{new}(t)$ denotes
the overall rate of infection to person $n$ from the new strain
and $\lambda_n^{old}(t)$ denotes the same quantity from the old strain,
so that 
\[
\lambda_n(t)= \lambda_n^{new}(t) + \lambda_n^{old}(t).
\]
Then, this person is assigned a new strain with probability
\[
\frac{\lambda_n^{new}(t)}{\lambda_n(t)}.
\]
and old strain with the remaining probability.
Else, the algorithm proceeds as before.

\section{October report results}
 \label{section:oct_res} 

\begin{figure}
      \centering
     \includegraphics[width=\linewidth]{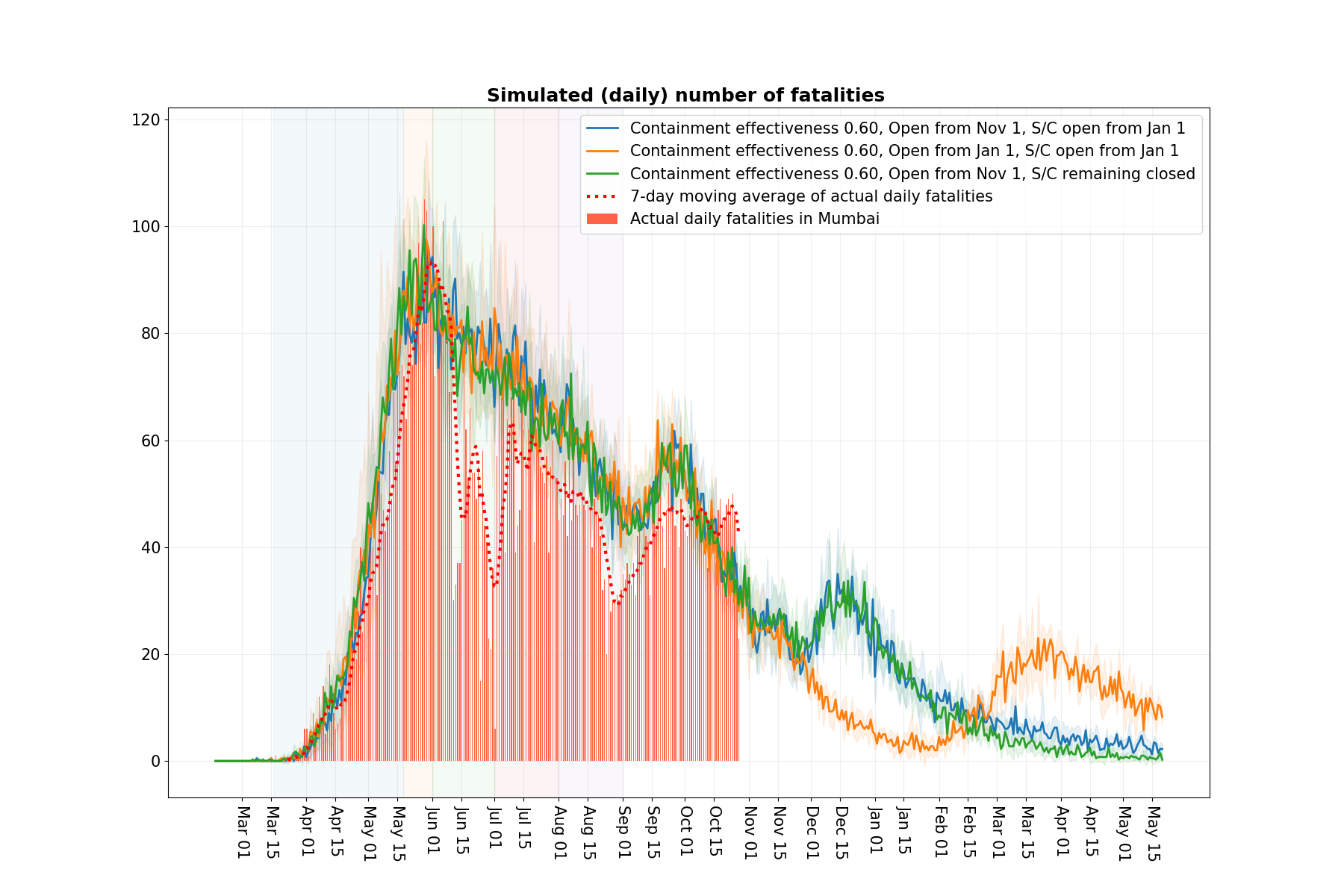}

      \caption{\small   Simulated number of  daily fatalities projections for Nov, Dec, Jan
 under the workplace opening schedule 
     5\% attendance, May 18
to May 31st, 15\% attendance in June, 25\% in July,
33\% in August, 50\% in September and October and fully open November onwards with School/Colleges opening from January 1. This schedule
is overlaid with scenarios of 1) workplace attendance of 100\% and school/colleges opening from January 1, 2) Workplace fully open from November 1 and school/colleges remaining closed. 
All the scenarios include the three festival relaxations.
} \label{open_figure_fatalities}
 
     \includegraphics[width=\linewidth]{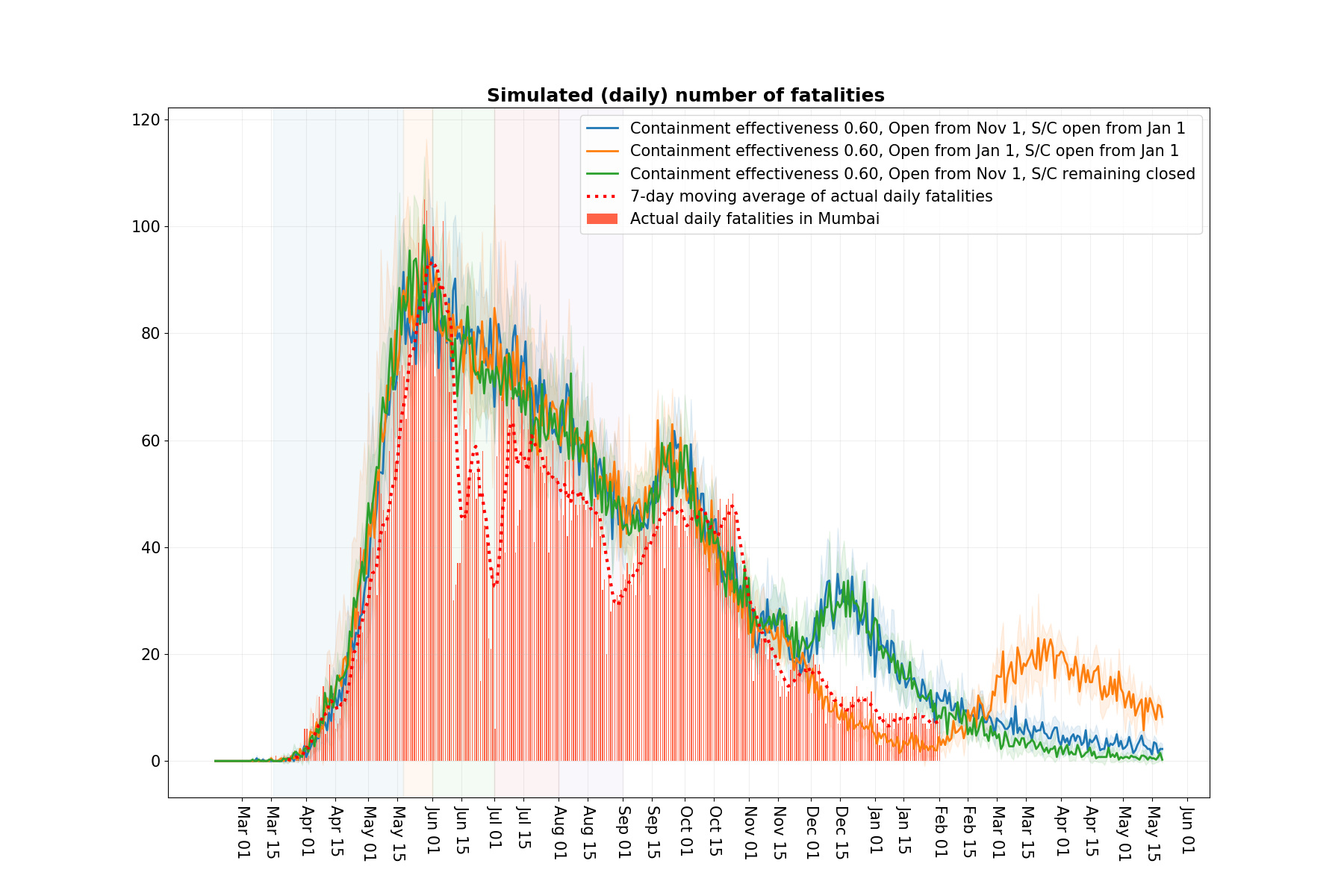}

      \caption{\small Figure~\ref{open_figure_fatalities} specifications with observed fatality data till January end.
} \label{open_figure_fatalities_real}
  \end{figure}

Figure~\ref{open_figure_fatalities} reproduces the fatality projections from our October Report
\cite{October_report_2020} where the scenario of the economy opening up on November 1, green curve, and the economy opening up
on January 1, orange curve,  are considered.
Figure~\ref{open_figure_fatalities_real} shows the same curve with the observed fatality numbers updated till February 1.
Since opening up on January 1 does not impact the fatalities till the end of January, the projections are valid till that time although
the major opening up in Mumbai happened on February 1. 
While the projections hold quite well till mid-December, 
Figure~\ref{open_figure_fatalities_real} shows that they underestimate thereafter. 
As the violet curve in Figure~\ref{daily_deaths_gen}   shows, this is significantly corrected once we assume increased laxity in population in December and January. 
In fact, that does a pretty good job of explaining fatalities right up to the first week of March. 
Of course, as we mentioned earlier, the subsequent increase in cases and fatalities
were difficult to explain simply by assuming laxity in the population, and are best explained by assuming
highly infective variants.

  \pagebreak
  
\section{Appendix}
\label{section:append}

 The contact rates used for experiments (see Fig \ref{fig:beta-values}) are same as in our earlier report~\cite{October_report_2020}.

\begin{figure}[h]
  \begin{center}
    \small
 \begin{tabular}{|ccc|}
      \hline
      {\bf Interaction space} & {\bf Comment}& {\bf $\beta$ value}\\
      \hline
      Home & (calibrated) & 1.93651 \\
      Workplace & (calibrated) & 0.26862\\
      Community & (calibrated) &  0.02152\\
      School & $2 \cdot \beta_{\text{workplace}}$ & 0.53723\\
      Project & $9 \cdot \beta_{\text{workplace}}$ & 2.41758\\
      Class & $9 \cdot \beta_{\text{school}}$ & 4.83507\\
      Neighbourhood & $9 \cdot \beta_{\text{community}}$ & 0.19368\\
      Close friends & $9 \cdot \beta_{\text{community}}$ & 0.19368\\
      \hline
    \end{tabular}
  \end{center}
  \caption{Interaction spaces, subnetworks and  contact rates  }
  \label{fig:beta-values}
  \end{figure}
  
\begin{figure}[h]
      \centering
     \includegraphics[width=\linewidth]{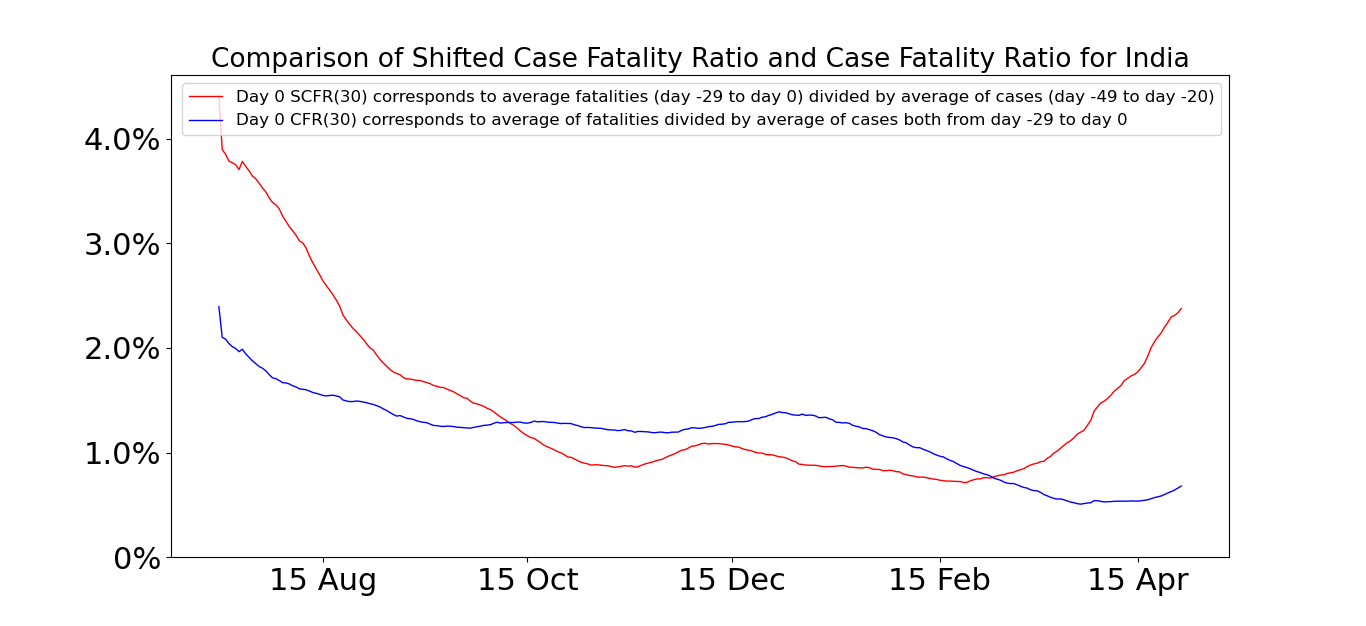}
      \caption{Comparison of Shifted Case Fatality Rate (Red curve)  and Case Fatality Rate (Blue curve) for India. Given 
      the current level of infection spread in India, clearly SCFR is a better measure of disease severity compared to CFR} \label{India_CFR_SCFR}

  \end{figure}

\begin{figure}
      \centering
     \includegraphics[width=\linewidth]{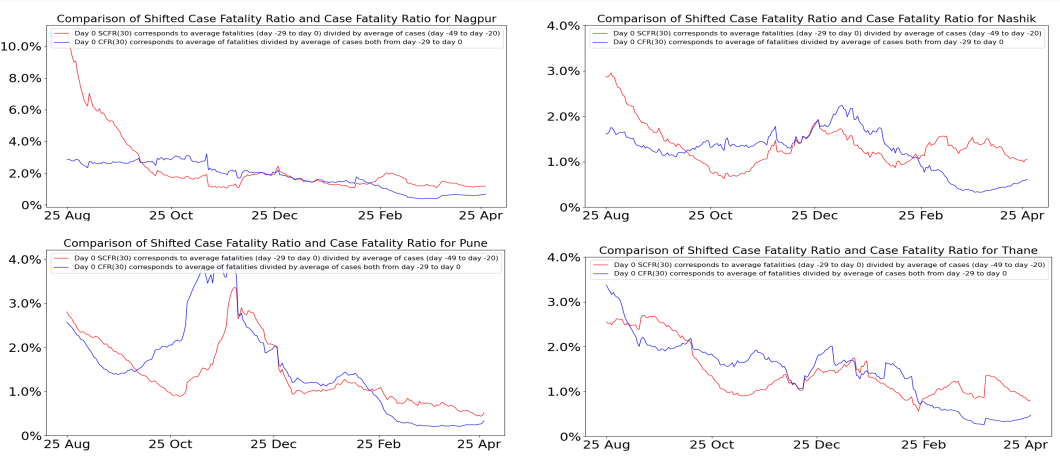}
      \caption{Comparison of Shifted Case Fatality Rate (Red curve)  and Case Fatality Rate (Blue curve) for different districts of Maharashtra (Nagpur, Nashik, Pune and Thane). 
      Nagpur has a SCFR nearing 2, while all the other districts are around 1, with Pune having the smallest SCFR value. 
      } \label{mh}

  \end{figure}

\begin{figure}
      \centering
     \includegraphics[width=\linewidth]{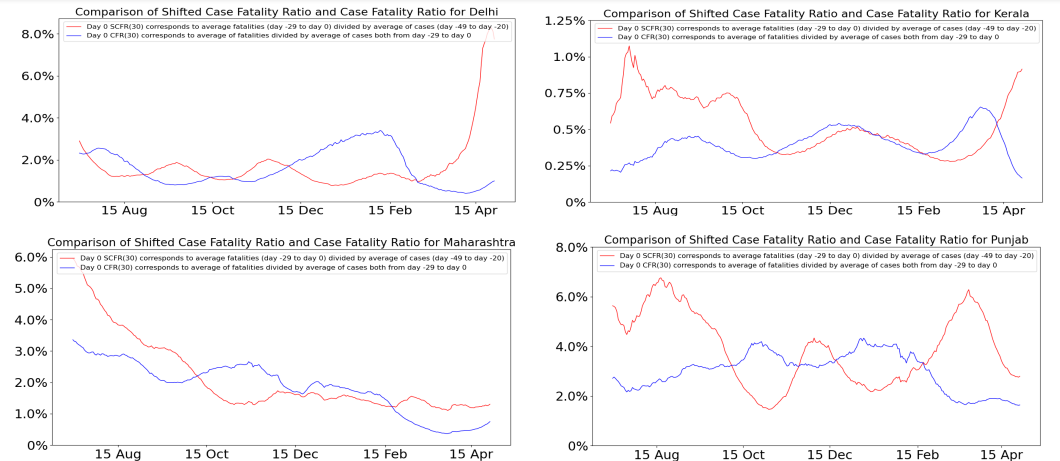}
      \caption{Comparison of Shifted Case Fatality Rate (Red curve) and Case Fatality Rate (Blue curve) for different states of India ( Delhi, Kerala, Maharashtra and Punjab ). } \label{india_1}

  \end{figure}

\begin{figure}
      \centering
     \includegraphics[width=\linewidth]{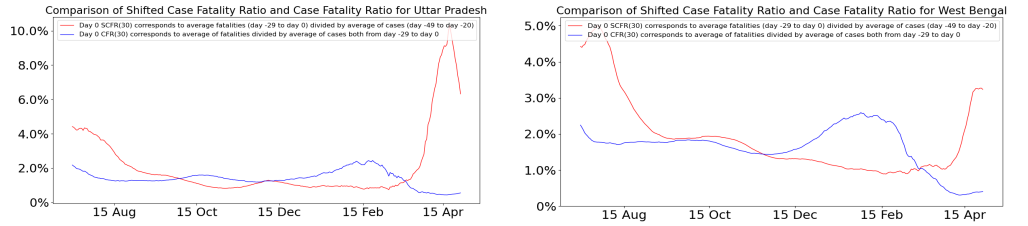}
      \caption{Comparison of Shifted Case Fatality Rate (Red curve)  and Case Fatality Rate (Blue curve) for different states of India (Uttar Pradesh and West Bengal). } \label{india_2}

  \end{figure}

 Following are the
SCFR graphs for India (Fig \ref{India_CFR_SCFR}), for other districts in Maharashtra (Fig \ref{mh}) as well as for some of the states (Fig \ref{india_1}, \ref{india_2})
with the largest number of people infected in the second wave.
  
\pagebreak

We refer the reader to our October Report \cite{October_report_2020}, Page 5 for additional caveats associated with this report.

%\pagebreak[]

\section*{Acknowledgments}

We thank our colleagues Prahladh Harsha, Ramprasad Saptharishi and   Piyush Srivastava for many useful suggestions that helped our analysis.  
We thank them as well as our IISc collaborators
R. Sundaresan,  P. Patil, N. Rathod, A. Sarath, S. Sriram, and N. Vaidhiyan for their tireless efforts
in developing the IISc-TIFR Simulation model \cite{City_Simulator_IISc_TIFR_2020}
and their key role
in our earliers report on  Mumbai. We thank IDFC Institute for sponsoring Daksh Mittal's work with the TIFR COVID-19 City-Scale Simulation Team.

We thank Mrs. Ashwini Bhide, AMC, MCGM for her insights and for her crucial data inputs. 
We also thank Shri Saurabh Vijay, Secretary, Higher \& Technical Education Department, Government of Maharashtra for his insights and data inputs. 

We  acknowledge the support of A.T.E. Chandra Foundation for this research.
We further  acknowledge the support of the Department of Atomic Energy, Government of India, to TIFR under project no. 12-R\&D-TFR-5.01-0500.

%\bibliographystyle{IEEEtran}
%{
%\bibliography{IEEEabrv,Mumbai_August}$}

\end{document}